\documentclass{SciPost}

\hypersetup{
    colorlinks,
    linkcolor={red!50!black},
    citecolor={blue!50!black},
    urlcolor={blue!80!black}
}

\usepackage[bitstream-charter]{mathdesign}
\DeclareSymbolFont{usualmathcal}{OMS}{cmsy}{m}{n}
\DeclareSymbolFontAlphabet{\mathcal}{usualmathcal}

\fancypagestyle{SPstyle}{
\fancyhf{}
\lhead{\colorbox{scipostblue}{\bf \color{white} ~SciPost Physics Codebases }}
\rhead{{\bf \color{scipostdeepblue} ~Submission }}

\fancyfoot[C]{\textbf{\thepage}}
}

\usepackage{listings} 
\usepackage{relsize} 
\usepackage{tcolorbox} 

\definecolor{halfgray}{gray}{0.55}
\definecolor{pyframe}{RGB}{207, 207, 207}
\definecolor{pybackground}{rgb}{0.95,0.95,0.95}
\definecolor{pypurple}{RGB}{170, 34, 255}

\definecolor{pykeyword}{HTML}{2A9D8F}
\definecolor{pyblue}{HTML}{4062BB}
\definecolor{ipython_red}{HTML}{52489C}
\definecolor{pycyan}{HTML}{E76F51}

\colorlet{outputcolor}{ipython_red!90!black}
\lstdefinelanguage{output}{
    basicstyle=\ttfamily\smaller[0.5]\color{outputcolor},
    identifierstyle=\ttfamily\color{outputcolor},
    numbers=none
}

\lstdefinelanguage{Python}{
    morekeywords={access,and,as,assert,break,class,continue,def,del,elif,else,except,exec,finally,for,from,global,if,import,in,is,lambda,not,or,pass,print,raise,return,try,while},
    morekeywords=[2]{abs,all,any,basestring,bin,bool,bytearray,callable,chr,classmethod,cmp,compile,complex,delattr,dict,dir,divmod,enumerate,eval,execfile,file,filter,float,format,frozenset,getattr,globals,hasattr,hash,help,hex,id,input,int,isinstance,issubclass,iter,len,list,locals,long,map,max,memoryview,min,next,object,oct,open,ord,pow,property,range,raw_input,reduce,reload,repr,reversed,round,set,setattr,slice,sorted,staticmethod,str,sum,super,tuple,type,unichr,unicode,vars,xrange,zip,buffer,coerce,intern,True,False,self},
    sensitive=true,
    morecomment=[l]\#,
    morestring=[b]',
    morestring=[b]",
    morestring=[s]{'''}{'''},
    morestring=[s]{"""}{"""},
    morestring=[s]{r'}{'},
    morestring=[s]{r"}{"},
    morestring=[s]{r'''}{'''},
    morestring=[s]{r"""}{"""},
    morestring=[s]{u'}{'},
    morestring=[s]{u"}{"},
    morestring=[s]{u'''}{'''},
    morestring=[s]{u"""}{"""},
    literate=
    {á}{{\'a}}1 {é}{{\'e}}1 {í}{{\'i}}1 {ó}{{\'o}}1 {ú}{{\'u}}1
    {Á}{{\'A}}1 {É}{{\'E}}1 {Í}{{\'I}}1 {Ó}{{\'O}}1 {Ú}{{\'U}}1
    {à}{{\`a}}1 {è}{{\`e}}1 {ì}{{\`i}}1 {ò}{{\`o}}1 {ù}{{\`u}}1
    {À}{{\`A}}1 {È}{{\'E}}1 {Ì}{{\`I}}1 {Ò}{{\`O}}1 {Ù}{{\`U}}1
    {ä}{{\"a}}1 {ë}{{\"e}}1 {ï}{{\"i}}1 {ö}{{\"o}}1 {ü}{{\"u}}1
    {Ä}{{\"A}}1 {Ë}{{\"E}}1 {Ï}{{\"I}}1 {Ö}{{\"O}}1 {Ü}{{\"U}}1
    {â}{{\^a}}1 {ê}{{\^e}}1 {î}{{\^i}}1 {ô}{{\^o}}1 {û}{{\^u}}1
    {Â}{{\^A}}1 {Ê}{{\^E}}1 {Î}{{\^I}}1 {Ô}{{\^O}}1 {Û}{{\^U}}1
    {œ}{{\oe}}1 {Œ}{{\OE}}1 {æ}{{\ae}}1 {Æ}{{\AE}}1 {ß}{{\ss}}1
    {ç}{{\c c}}1 {Ç}{{\c C}}1 {ø}{{\o}}1 {å}{{\r a}}1 {Å}{{\r A}}1
    {€}{{\EUR}}1 {£}{{\pounds}}1
    {^}{{{\color{pypurple}\^{}}}}1
    {=}{{{\color{pypurple}=}}}1
    {+}{{{\color{pypurple}+}}}1
    {*}{{{\color{pypurple}$^\ast$}}}1
    {/}{{{\color{pypurple}/}}}1
    {+=}{{{+=}}}1
    {-=}{{{-=}}}1
    {*=}{{{$^\ast$=}}}1
    {/=}{{{/=}}}1,
    literate=
    *{-}{{{\color{pypurple}-}}}1
     {?}{{{\color{pypurple}?}}}1,
    identifierstyle=\color{black}\ttfamily,
    commentstyle=\color{pycyan}\ttfamily,
    stringstyle=\color{ipython_red}\ttfamily,
    showstringspaces = false,
    keepspaces=true,
    showspaces=false,
    showstringspaces=false,
    rulecolor=\color{pyframe},
    frame=single,
    frameround={t}{t}{t}{t},
    framexleftmargin=6mm,
    numbers=left,
    numberstyle=\tiny\color{halfgray},
    backgroundcolor=\color{pybackground},
    basicstyle=\ttfamily\smaller[0.5],
    keywordstyle=\color{pykeyword}\ttfamily,
}

\newcommand{\code}[1]{\tcbox[on line,colback=pybackground,colframe=white,boxsep=3pt,left=0pt,right=0pt,top=0pt,bottom=0pt]{\lstinline[language=python,breaklines=true,breakatwhitespace=true]|#1|}} 

\usepackage{float}
\newfloat{listing}{tbp}{lol}
\floatname{listing}{Listing}

\usepackage{cleveref}
\crefname{listing}{listing}{listings}
\crefname{section}{Sec.}{Secs.} 
\Crefname{section}{Section}{Sections}
\crefname{figure}{Fig.}{Figs.} 
\Crefname{figure}{Figure}{Figures}
\crefformat{footnote}{#2\footnotemark[#1]#3}

\usepackage{braket}
\usepackage{bbm} 
\usepackage{amsthm}

\usepackage{lipsum}

\usepackage{enumitem}
\usepackage{algorithm}
\usepackage{algpseudocode}

\newcommand{\unity}{\mathbbm{1}}

\newcommand{\vac}{\mathrm{vac}}
\newcommand{\T}{\mathrm{T}}

\let\Im\relax

\DeclareMathOperator{\Im}{Im}
\DeclareMathOperator{\tr}{tr}
\DeclareMathOperator{\pf}{pf}
\DeclareMathOperator{\diag}{diag}

\begin{document}

\pagestyle{SPstyle}

\begin{center}{\Large \textbf{\color{scipostdeepblue}{
TeMFpy: a Python library for converting fermionic mean-field states into tensor networks\\
}}}\end{center}

\begin{center}\textbf{
Simon H. Hille\textsuperscript{$\star$},
Attila Szabó\textsuperscript{$\dagger$}
}\end{center}

\begin{center}
Department of Physics, University of Zürich, Zürich, Switzerland
\\[\baselineskip]
$\star$ \href{mailto:simon.hille@physik.uzh.ch}{\small simon.hille@physik.uzh.ch}\,,\quad
$\dagger$ \href{mailto:attila.szabo@physik.uzh.ch}{\small attila.szabo@physik.uzh.ch}
\end{center}

\section*{\color{scipostdeepblue}{Abstract}}
\textbf{\boldmath{%
We introduce TeMFpy, a Python library for converting fermionic mean-field states to finite or infinite matrix product state (MPS) form.
TeMFpy includes new, efficient, and easy-to-understand algorithms for both Slater determinants and Pfaffian states. Together with Gutzwiller projection, these also allow the user to build variational wave functions for various strongly correlated electron systems, such as quantum spin liquids.
We present all implemented algorithms in detail and describe how they can be accessed through TeMFpy, including full example workflows.
TeMFpy is built on top of TeNPy and, therefore, integrates seamlessly with existing MPS-based algorithms.
}}

\vspace{\baselineskip}

\noindent\textcolor{white!90!black}{%
\fbox{\parbox{0.975\linewidth}{%
\textcolor{white!40!black}{\begin{tabular}{lr}%
  \begin{minipage}{0.6\textwidth}%
    {\small Copyright attribution to authors. \newline
    This work is a submission to SciPost Physics Codebases. \newline
    License information to appear upon publication. \newline
    Publication information to appear upon publication.}
  \end{minipage} & \begin{minipage}{0.4\textwidth}
    {\small Received Date \newline Accepted Date \newline Published Date}%
  \end{minipage}
\end{tabular}}
}}
}


\vspace{10pt}
\noindent\rule{\textwidth}{1pt}
\tableofcontents
\noindent\rule{\textwidth}{1pt}
\vspace{10pt}


\section{Introduction}
\label{sec:intro}

Fermionic mean-field (or Gaussian~\cite{Bravyi2005FermionicGaussian}) states are a class of variational wave functions of central importance.
All eigenstates of non-interacting fermionic Hamiltonians are Gaussian.
As such, they form the starting point of such mean-field approaches as Hartree--Fock theory and the BCS theory of superconductivity, which often provide an excellent account of strongly interacting quantum systems as well.
Furthermore, ``dressing'' mean-field states with, e.g., Gutzwiller and Jastrow factors gives rise to even more powerful variational ansätze, which are widely deployed in variational quantum Monte Carlo (VMC)~\cite{Becca2017QuantumSystems}.
Through parton constructions and Gutzwiller projection, fermionic mean-field states also form the basis of the most widely used variational ansätze for quantum spin liquids~\cite{Becca2017QuantumSystems,Wen2002QuantumLiquids,Wen2007QuantumElectrons,Savary2017QuantumReview}.

Mean-field states are, however, a special class of wave function and cannot fully account for interaction-induced phenomena.
A more general approach is offered by tensor networks, which can represent mean-field and interacting quantum states equally well, with an expressive power systematically improvable through the number of variational parameters.
Matrix product states (MPS), designed for one-dimensional systems, are particularly useful
on account of such powerful ground-state optimisation algorithms as the density matrix renormalisation group (DMRG)~\cite{Schollwock2011DMRG} and variational uniform MPS (VUMPS)~\cite{Vanderstraeten2019vumps}.

Given the individual success of these two approaches, it is desirable to combine the two.
In particular, DMRG and VUMPS are typically initialised with a random MPS, which leaves them prone to converging to local minima, which potentially display qualitatively different physics to the true ground state.
(Dressed) mean-field states (from, e.g., Hartree--Fock theory or VMC) are a much closer starting point, stabilising and reducing the computational cost of MPS optimisation~\cite{Jin2021GutzwillerDMRG,Jin2025GutzwillerDMRGReview}.
By initialising with different ansätze, we may also compare different mean-field theories as approximate descriptions of the physical system~\cite{Sun2024KagomeCSL}.
Using mean-field states as initial guesses, however, requires efficient algorithms to convert fermionic mean-field wave functions to an MPS form.

Furthermore, MPS representations give ready access to such quantities as the entanglement spectrum and entropy, which are difficult to obtain from VMC but often carry detailed information of exotic physics, such as topological order and the associated edge physics~\cite{LiHaldane2008,Qi2012EntanglementSpectrum}.
This is a particularly appealing prospect for Gutzwiller-projected mean-field states, which are common ansatz wave functions for quantum spin liquids~\cite{Becca2017QuantumSystems,Wen2002QuantumLiquids,Wen2007QuantumElectrons,Savary2017QuantumReview}, fractional quantum Hall states~\cite{Jain1989Parton,Lu2012PartonFCI,McGreevy2012PartonFCI}, and other topological orders~\cite{Li2025PartonSUnk}, as the Gutzwiller projection is a local operation, so it can easily be applied to real-space MPS representations of mean-field states.

As first noted in Ref.~\cite{Petrica2021MF_MPS}, Slater determinants can conveniently be converted to MPS form, since their Schmidt vectors are themselves Slater determinants~\cite{Peschel2003GaussianReducedDM,Cheong2004GaussianSchmidt,Peschel2012GaussianSchmidtReview}
and so entries of the canonical MPS tensors are overlaps of Slater determinants, which can be computed efficiently.
Since then, this idea has been expanded to Pfaffian (or Bogoliubov mean-field) states~\cite{Jin2022BogoliubovMPS} and successfully used to study Gutzwiller-projected mean-field states~\cite{Wu2022SU22} and as starting points to DMRG~\cite{Jin2021GutzwillerDMRG,Jin2025GutzwillerDMRGReview,Sun2024KagomeCSL}.

In this paper, we introduce a new open-source Python library called \textbf{TeMFpy} for efficiently converting fermionic mean-field states to MPS form.
TeMFpy provides both an easy-to-use interface for the key functionality of representing Slater determinants and Pfaffian states as either finite or infinite MPS and a number of well-defined building blocks for advanced users to build new algorithms.
It is built on top of TeNPy~\cite{tenpy}, so that the resulting MPS representations can be used directly for further calculations with, e.g., DMRG.
We hope that TeMFpy will open up the power of combining mean-field, VMC, and tensor-network approaches to a wide community of users.

\subsection{Outline}

The core functionality, constructing finite MPS representations of fermionic mean-field states, specified through either a (Nambu) mean-field Hamiltonian or a (Nambu) correlation matrix, is described in \cref{sec: mps}.
In particular, \Cref{sec: mps user guide} provides an overview of the interface and a guide to basic usage.

We introduce our new algorithms for computing MPS tensor entries for Slater determinants and Pfaffian states in detail in \cref{sec: slater,sec: pfaffian}, respectively.
In both cases, we exploit the fact that the leading Schmidt vectors that we keep only differ in the occupation of a few ``entangled'' orbitals~\cite{Liu2025EfficientGaussianMPS,Li2025EfficientGaussianMPS}, so their overlaps can be reduced to a determinant or a Pfaffian defined only on these orbitals.
In a typical Schmidt decomposition, the $\chi$ leading Schmidt vectors can be captured through $n_\mathrm{ent} = O(\log\chi)$ entangled orbitals, so a single MPS tensor entry can be computed in $O(n_\mathrm{ent}^3) = O(\log^3\chi)$, rather than $O(L^3)$, time, a dramatic speedup.

\Cref{sec: imps} discusses TeMFpy's functionality for computing infinite MPS, based on ideas outlined in Ref.~\cite{Petrica2021MF_MPS}.  We provide both a generic implementation able to convert any finite MPS into infinite ones, as well as specialised algorithms for mean-field states in \cref{sec: imps mean field}, which bypass the need to construct finite MPS for large systems.
Finally, we implement Gutzwiller projection schemes in \cref{sec: gutzwiller}.

\subsection{Installing TeMFpy}

TeMFpy is written in pure Python (the current version 0.3.3 requires at least Python 3.10 and TeNPy 1.1.0) and is largely platform-agnostic.\footnote{Installing the \code{pfapack} dependency (needed for handling Pfaffian states) for Windows is, however, not trivial, see \href{https://pfapack.readthedocs.io/en/latest/\#usage}{here} for details. You can also bypass this issue using the \href{https://ubuntu.com/desktop/wsl}{Windows Subsystem for Linux}.}
It is available from PyPI using the command
\begin{lstlisting}
pip install --upgrade temfpy
\end{lstlisting}
We will also make it available on CondaForge soon.
The development version is available at \url{https://github.com/temfpy/temfpy}, where we also welcome issue reports, suggestions, and code contributions.
A detailed API reference, examples, and tutorials are available online at \url{https://temfpy.github.io/temfpy}.

\section{Converting mean-field states to MPS}
\label{sec: mps}

Writing any quantum state as an MPS (in canonical form) relies on computing its Schmidt decomposition,
\begin{equation}
    \ket\psi = \sum_\alpha \lambda_\alpha^{(i)} \ket{L_\alpha^{(i)}} \otimes_g \ket{R_\alpha^{(i)}},
    \label{eq: schmidt decomp general}
\end{equation}
where $\lambda_\alpha^{(i)}$ are the Schmidt values across entanglement cut $i$, while  $\ket{L_\alpha^{(i)}}$ and $\ket{R_\alpha^{(i)}}$ are the corresponding Schmidt vectors defined on the sites left and right to the same cut.%
\footnote{In this paper, we always deal with Hilbert spaces of fermions, so the standard tensor product is replaced by the fermionic ($\mathbb{Z}_2$ graded) one, $\otimes_g$~\cite{Bultinck2017FermionicMPS}. This distinction will be particularly important when we come to Pfaffian states.}
Now, the entries of the left-canonical MPS tensor for a given site are given by the overlaps of the left Schmidt vectors for the entanglement cuts on either side of it~\cite{Schollwock2011DMRG}:
\begin{subequations}
\label{eq: MPS tensor overlap}
\begin{equation}
        A^{n_i}_{\alpha\beta} =
        \left(\langle n_i | \otimes_g \langle L^{(i-1)}_{\alpha}|\right)\;
        |L^{(i)}_{\beta} \rangle,
\end{equation}
while right-canonical tensors are given by overlaps of right Schmidt vectors:
\begin{equation}
        B^{n_i}_{\beta\alpha} =
        \left(\langle R^{(i)}_\alpha| \otimes_g \langle n_i|\right)\;
        |R^{(i-1)}_\beta \rangle.
\end{equation}
\end{subequations}

For a generic state, Eq.~\eqref{eq: MPS tensor overlap} is impractical to use directly.
However, the eigenstates of fermionic mean-field Hamiltonians~\cite{Blaizot1985} of the form
\begin{equation}
    H = \sum_{ij} \left[ t_{ij} f^\dagger_i f_j  + (\Delta_{ij} f^\dagger_i f^\dagger_j + \mathrm{h.c.}) \right]
\end{equation}
are fermionic Gaussian states~\cite{Bravyi2005FermionicGaussian}, that is, Slater determinants (if there are no pairing terms $\Delta_{ij}$)
or Pfaffian (i.e., Bogoliubov mean-field) states.
The Schmidt vectors of such states are in turn also Slater determinants or Pfaffian states and the (generalised) orbitals that define them can be obtained from diagonalising their single-particle correlation matrices~\cite{Peschel2003GaussianReducedDM,Cheong2004GaussianSchmidt,Peschel2012GaussianSchmidtReview}.
The entries of the MPS tensor~\eqref{eq: MPS tensor overlap} thus reduce to overlaps of fermionic Gaussian states, which can be computed efficiently as determinants or Pfaffians of the overlaps of individual orbitals.

Furthermore, as the leading Schmidt vectors only differ in the occupation of a few of these (generalised) orbitals~\cite{Liu2025EfficientGaussianMPS,Li2025EfficientGaussianMPS}, we can apply linear-algebra manipulations to these determinants and Pfaffians to reduce the computational complexity of computing each MPS tensor from $O(\chi^2 N^3)$~\cite{Petrica2021MF_MPS} to $O(N^3 + \chi^2\log^3\chi)$, where $N$ is the size of the system and $\chi$ is the bond dimension of the tensor.
Our algorithms thus differ significantly from those of Ref.~\cite{Liu2025EfficientGaussianMPS,Li2025EfficientGaussianMPS}, as we do not rely on introducing an array of virtual fermion degrees of freedom and effectively compress several intermediate steps in their approach into one.
We describe our method in detail in \cref{sec: slater fast overlap,sec: pfaffian fast overlap}.

\paragraph{Warning for finite bond dimension.}

In the limit $\chi\to\infty$, our method is exact as the MPS tensors are built from the overlaps of the exact Schmidt vectors and none of the wave function has to be truncated. 
In this ideal case, the final MPS is normalised, i.e., $\braket{\mathrm{MPS}|\mathrm{MPS}}=1.0$.
For finite bond dimensions, however, some of the wave function must be truncated at every entanglement cut. 
This both lowers the norm $\braket{\mathrm{MPS}|\mathrm{MPS}}$ from 1.0 and makes the MPS tensors slightly noncanonical.
This issue is particularly important for cylinder MPS simulations of two-dimensional systems, where truncation errors are substantial even for large bond dimensions.
Truncation error estimates for each MPS tensor are printed during the MPS conversion if logging is turned on using \code{temfpy.setup_logging()}.
The overall conversion error can be estimated from the deviation of the norm of the resulting MPS from 1.0.

\subsection{How to use the TeMFpy implementation}
\label{sec: mps user guide}

TeMFpy implements the calculations outlined above for both Slater determinants (module \code{temfpy.slater}) and Pfaffian states (module \code{temfpy.pfaffian}). Both follow the same general structure and calling sequences:
\begin{enumerate}
    \item \code{SchmidtModes.from_correlation_matrix()} computes (generalised) orbitals that build up the Schmidt vectors by diagonalising submatrices of the single-particle correlation matrix. The resulting orbital information is kept in \code{SchmidtModes} objects.
    \item \code{SchmidtVectors.from_schmidt_modes()} finds the combinations of these orbitals that correspond to the largest Schmidt values.%
    \footnote{Each Schmidt value $\lambda_\alpha^{(i)}$ is given by a product of eigenvalues from the previous diagonalisation, so finding the leading ones has the same complexity as finding the subsets of a set with the highest products. See Ref.~\cite{LowestSums} and \Cref{app: highest schmidt values} for details of how we solve this latter problem.}
    These combinations and the corresponding Schmidt values are stored in \code{SchmidtVectors} objects, which can be thought of as compressed representations of the leading Schmidt vectors on a given entanglement cut.

    For convenience, we also provide \code{SchmidtVectors.from_correlation_matrix()} that combines the functionality of the two functions above.
    \item \code{MPSTensorData.from_schmidt_vectors()} then takes two \code{SchmidtVectors} objects corresponding to the entanglement cuts on either side of a site, precomputes quantities related to orbital overlaps (see \cref{sec: slater fast overlap,sec: pfaffian fast overlap} for details), and stores them in an \code{MPSTensorData} object, which is effectively a compressed description of the MPS tensor on the given site.
    
    Finally, \code{MPSTensorData.to_npc_array()} converts this compressed format into an explicit MPS tensor.
\end{enumerate}
In addition, the following convenience functions are provided:
\begin{itemize}
    \item \code{correlation_matrix()} computes the ground-state correlation matrix of a given gapped\footnote{\label{fn: gapped}If some modes are gapless, there is no unique ground state whose correlation matrix could be returned. It is up to the user to fix a choice of ground state and construct its correlation matrix.} (tight-binding or Nambu) Hamiltonian.
    \item \code{C_to_MPS()} performs steps 1--3 for every site of the system to convert a correlation matrix directly into a TeNPy MPS object in mixed canonical form.
    \item \code{C_to_iMPS()} uses two correlation matrices of a gapped system that differ by one repeating unit cell to construct an infinite MPS representation of the ground state in the thermodynamic limit. This function and its usage is described in more detail in \cref{sec: imps mean field}.
    \item \code{H_to_MPS()} and \code{H_to_iMPS()} combine the functions above: starting from a gapped\cref{fn: gapped} Hamiltonian, it first computes the ground-state correlation matrices, which are then used to construct the MPS.
\end{itemize}
The basic usage of these functions (except \code{C_to_iMPS} and \code{H_to_iMPS}) and objects is demonstrated in \Cref{listing: basic usage}.

\begin{listing}[!t]
\begin{lstlisting}
import numpy as np
from temfpy import slater

# Hamiltonian, in this case a 10-site tight-binding chain
H = -(np.eye(10, k=1) + np.eye(10, k=-1))

# Ground-state correlation matrix
C, _ = slater.correlation_matrix(H) # also returns particle number

# Control parameters for MPS truncation
trunc_par = {
    "chi_max": 100, # maximum bond dimension
    "svd_min": 1e-6, # lowest Schmidt value kept (default: 1e-6)
    "degeneracy_tol": 1e-12, # largest relative difference where Schmidt values are considered degenerate (default: 1e-12)
}

# Orbitals that make up the Schmidt vectors (cut between sites 4 and 5)
modes = slater.SchmidtModes.from_correlation_matrix(C, 5, trunc_par)

# Schmidt vectors in terms of these orbitals
schmidt = slater.SchmidtVectors.from_correlation_matrix(C, 5, trunc_par)
schmidt = slater.SchmidtVectors.from_schmidt_modes(modes, trunc_par)

# MPS tensor for site 5
# first the Schmidt vectors for the cut to the right
schmidt_R = slater.SchmidtVectors.from_correlation_matrix(C, 6, trunc_par)
# left canonical
A = slater.MPSTensorData.from_schmidt_vectors(schmidt, schmidt_R, "left")
A = A.to_npc_array() # explicit MPS tensor as a TeNPy array
# right canonical
B = slater.MPSTensorData.from_schmidt_vectors(schmidt_R, schmidt, "right")
B = B.to_npc_array() # explicit MPS tensor as a TeNPy array

# Build the full MPS from the correlation matrix
mps = slater.C_to_MPS(C, trunc_par)
# or straight from the Hamiltonian
mps = slater.H_to_MPS(H, trunc_par)
\end{lstlisting}
    \caption{Basic usage of the \code{temfpy.slater} module. The calling sequence of the corresponding functions in \code{temfpy.pfaffian} is identical, except that one must specify the full Nambu Hamiltonian/correlation matrix in the input (cf.\ \cref{sec: pfaffian basis}).}
    \label{listing: basic usage}
\end{listing}

\subsection{Slater determinants}
\label{sec: slater}

\subsubsection{Schmidt decomposition}
Given the correlation matrix $C_{ij} := \braket{c_j^\dagger c_i}$ for a Slater determinant, we start by diagonalizing the correlation matrix restricted to subsystem $A$:
\begin{equation}
    C^{AA}_{ij} := \braket{c^\dagger_j c_i}_{i,j \in A} = U_A\Lambda_AU_A^\dagger.
\end{equation}
The eigenvectors $U_A$ define a new set of orbitals
\begin{equation}
    d^\dagger_{A,\alpha} = \sum_{i \in A} (U_A)_{i,\alpha} c^\dagger_i.
\end{equation}
The corresponding eigenvalues $\lambda_{A,\alpha}$ are between~0 and~1 (cf.~\Cref{app: svd}):
If $\lambda_{A,\alpha}=1$ (0), $d^\dagger_{A,\alpha}$ is an occupied (empty) orbital of the overall Slater determinant, contained entirely within subsystem $A$.
Orbitals corresponding to eigenvalues $0 < \lambda_{A,\alpha} < 1$ are entangled with the rest of the system.

Consider now a chain split into left and right halves by an entanglement cut.
By diagonalizing the correlation matrices restricted to these subsystems, we obtain two sets of operators $d^\dagger_L$ and $d^\dagger_R$.
When transforming the correlation matrix into the combined basis of these, the entangled orbitals on either side form $2 \times 2$ blocks of the form
\begin{equation}
    \left(
    \begin{array}{cc}
        \braket{d_{L,\alpha}^\dagger \, d_{L,\alpha}} & \braket{d_{R,\alpha}^\dagger \, d_{L,\alpha}} \\
        \braket{d_{L,\alpha}^\dagger \, d_{R,\alpha}} & \braket{d_{R,\alpha}^\dagger \, d_{R,\alpha}}
    \end{array}
    \right)_{\alpha \in E}
    =
    \left(
    \begin{array}{cc}
        \lambda_\alpha & \sqrt{\lambda_\alpha(1-\lambda_\alpha)} \\
         \sqrt{\lambda_\alpha(1-\lambda_\alpha)} & 1 - \lambda_\alpha
    \end{array}
    \right),
\end{equation}
see \Cref{app: svd}.
Each of these blocks yields one occupied orbital in the overall Slater determinant, with creation operator
\begin{equation}
    b^\dagger_\alpha = \sqrt{\lambda_\alpha} d^\dagger_{L,\alpha} + \sqrt{1-\lambda_\alpha} d^\dagger_{R, \alpha}.
\end{equation}
Therefore, the Slater determinant can be written in terms of the occupied orbitals of the left ($O_L$) and right ($O_R$) subsystems and the entangled orbitals ($E$) shared between the two subsystems:
\begin{equation}
	\label{eq: generalSlaterDet}
	\ket{\psi} =
	\prod_{\mu \in O_L} d^\dagger_{L, \mu}
	\prod_{\alpha \in E} \left(\sqrt{\lambda_\alpha} d^\dagger_{L,\alpha} + \sqrt{1 - \lambda_\alpha} d^\dagger_{R,\alpha} \right)
	\prod_{\nu \in O_R} d^\dagger_{R,\nu} \ket{0}.
\end{equation}
Multiplying (\ref{eq: generalSlaterDet}) out shows that the Schmidt values are products with a term of either $\sqrt{\lambda_\alpha}$ or $\sqrt{1 - \lambda_\alpha}$ for each $\alpha\in E$.
The corresponding Schmidt vectors are themselves Slater determinants, with each entangled orbital occupied in either the left (if the corresponding factor in the Schmidt value is $\sqrt{\lambda_\alpha}$) or the right Schmidt vector (if the corresponding factor in the Schmidt value is $\sqrt{1 - \lambda_\alpha}$).

\subsubsection{Fast overlap computation}
\label{sec: slater fast overlap}

Evaluating~\eqref{eq: MPS tensor overlap} thus requires us to compute the overlaps of Slater determinants. (The on-site degree of freedom can be included in the ``bra'' Slater determinant as an additional site-localised orbital~\cite{Petrica2021MF_MPS}.)
It is well known that such overlaps are given by the determinant of the matrix of overlaps of the individual orbitals;
evaluating these determinants independently would yield the whole MPS tensor in $O(\chi^2N^3)$ time~\cite{Petrica2021MF_MPS}, where $\chi$ is the bond dimension of the MPS.

\begin{figure}[t]
    \centering
    \includegraphics{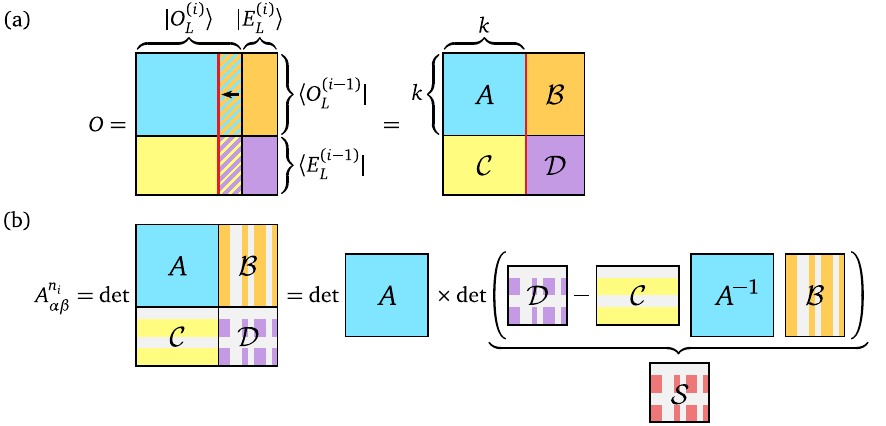}
    \caption{\textbf{(a)} In order to make the submatrix $A$ of $O$~\eqref{eq: bigO} that contains the overlaps of occupied orbitals, we may need to reclassify some occupied orbitals as entangled.
    The size $k$ of $A$ is the smaller of the number of occupied orbitals in the bra and the ket state.
    \textbf{(b)} The matrix $A$ appears in all overlap matrices~\eqref{eq: block determinant} that determine the elements of the left-canonical tensor $A^{n_i}_{\alpha \beta}$.  
    The remaining blocks are subsets of the rows/columns of the submatrices $\mathcal{B}, \mathcal{C},$ and $\mathcal{D}$ (gray lines indicate the rows/columns left out), 
    depending on the entangled orbitals in the respective bra (rows) and ket (columns) Schmidt vectors.
    To efficiently compute the tensor entries, we precompute $\det(A)$ and the matrix $\mathcal{S}$~\eqref{eq: S matrix} and select the appropriate rows and columns of $\mathcal{S}$ for each tensor element.
    }
    \label{fig: detOverlap}
\end{figure}

TeMFpy improves on this scaling by noting that the overlaps of the ``occupied'' orbitals in~\eqref{eq: generalSlaterDet} appear in all of these determinants,
i.e., the determinants can be written in the following block structure:%
\footnote{For the sake of simplicity, we follow the case of left Schmidt vectors in what follows. Right Schmidt vectors can be dealt with analogously.}
\begin{equation}
    A^{n_i}_{\alpha\beta} = \begin{vmatrix}
        \braket{O^{(i-1)}_L| O^{(i)}_L} & \braket{O^{(i-1)}_L| E^{(i)}_{L,\beta}} \\[1ex]
        \braket{E^{(i-1)}_{L,\alpha}| O^{(i)}_L} & \braket{E^{(i-1)}_{L,\alpha}| E^{(i)}_{L,\beta}}
    \end{vmatrix} =: \begin{vmatrix} A&B\\C&D \end{vmatrix},
    \label{eq: block determinant}
\end{equation}
where $\ket{O_L}$ and $\ket{E_\alpha}$ stand for the wave functions of the occupied orbitals and the entangled orbitals $d^\dagger_L$ contained in the Schmidt vector $\ket{L_\alpha}$, respectively. 
[We may reclassify a few occupied orbitals as entangled to make the block $A$ square, see \cref{fig: detOverlap}(a).]
The submatrix $A$ appears identically in each determinant, which motivates evaluating~\eqref{eq: block determinant} using the identity%
\footnote{For right Schmidt vectors, the entangled orbitals come first, so we use $|\cdot| = \det (D)\det(A-BD^{-1}C)$ instead.}
\begin{equation}
     \begin{vmatrix} A&B\\C&D \end{vmatrix} = \det (A)\det(D-CA^{-1}B).
     \label{eq: block matrix determinant formula}
\end{equation}
As illustrated in \cref{fig: detOverlap}(b), each entry $(D-CA^{-1}B)_{ab}$ depends only on the two entangled orbitals $d^{(i-1)}_{L,a}$ and $(d^{(i)}_{L,b})^\dagger$.
Therefore, they can be precomputed for every pair of entangled orbitals as the single matrix
\begin{align}
    \mathcal{S} &= \mathcal{D} - \mathcal{C}A^{-1}\mathcal{B}
    \label{eq: S matrix}\\
    O & := \begin{pmatrix}
        \braket{O^{(i-1)}_L| O^{(i)}_L} & \braket{O^{(i-1)}_L| E^{(i)}_{L}} \\[1ex]
        \braket{E^{(i-1)}_{L}| O^{(i)}_L} & \braket{E^{(i-1)}_{L}| E^{(i)}_{L}} \end{pmatrix} 
        := \begin{pmatrix} A&\mathcal{B}\\ \mathcal{C}&\mathcal{D} \end{pmatrix},
        \label{eq: bigO}
\end{align}
where $\ket{E_L^{(i-1)}}$ and $\ket{E_L^{(i)}}$ stand for all the entangled orbitals across the respective entanglement cuts.
This precomputation takes $O(N^3)$ time, after which each entry of the MPS tensor requires computing a determinant of typical size $O(\log\chi)$, resulting in an overall time cost of $O(N^3 + \chi^2\log^3\chi)$.
This matches the asymptotic complexity of~\cite{Liu2025EfficientGaussianMPS,Li2025EfficientGaussianMPS}, but by forgoing the construction of intermediate fermionic orbitals, we achieve equivalent results in the fewest possible steps, thus likely improving on wall-clock time.

\subsection{Pfaffian (Bogoliubov mean-field) states}
\label{sec: pfaffian}

\subsubsection{How to specify Hamiltonians and correlation matrices}
\label{sec: pfaffian basis}

From the user's point of view, the main complication of the Pfaffian case compared to Slater determinants is the appearance of anomalous correlations $\braket{cc}$, $\braket{c^\dagger c^\dagger}$, which need to be specified together with the number-conserving ones.
There are a number of conventions one can choose to specify the overall Nambu Hamiltonians and correlation matrices, of which TeMFpy supports two:

\paragraph{Complex-fermion basis.}

The ``basis'' of single-particle operators are the creation and annihilation operators $c_i^\dagger$ and $c_i$, listed as the row vector $(c_i^\dagger, c_i)$. The Nambu Hamiltonian matrix $H$ is interpreted as $\hat H = (c_i^\dagger, c_i) \;H\; (c_j^\dagger, c_j)^\dagger$, i.e., it is expected to contain the coefficients of
\begin{equation*}
    \begin{pmatrix}
        c_i^\dagger c_j & c_i^\dagger c_j^\dagger \\[2pt]
        c_i         c_j & c_i         c_j^\dagger 
    \end{pmatrix}
\end{equation*}
in this layout, while the correlation matrix is given by~\cite{Blaizot1985}
\begin{equation}
    C = \unity - \left\langle (c_i^\dagger, c_i)^\dagger\; (c_j^\dagger, c_j)\right\rangle = \begin{pmatrix}
            \langle c_j^\dagger c_i         \rangle & \langle c_j c_i         \rangle \\[2pt]
            \langle c_j^\dagger c_i^\dagger \rangle & \langle c_j c_i^\dagger \rangle
    \end{pmatrix}.
    \label{eq: nambu correlation matrix}
\end{equation}
In particular, TeMFpy expects that the matrices consist of $2\times2$ blocks containing the Hamiltonian terms or correlators for a given pair of sites, i.e., that \code{C[2*i : 2*i+2, 2*j : 2*j+2]} contain the correlators~\eqref{eq: nambu correlation matrix} for sites $i$ and $j$.

\paragraph{Majorana basis.}

Instead of the complex-fermion operators $c, c^\dagger$, we may also introduce the Majorana operators
\begin{align}
        \gamma_{2n} &= (c^\dagger_n + c_n) / \sqrt{2},&
        \gamma_{2n+1} &= i(c^\dagger_n - c_n) / \sqrt{2}
        \label{eq: majorana}
\end{align}
to define Nambu Hamiltonians and correlation matrices. Note that the normalisation of~\eqref{eq: majorana} differs from the usual one by a factor of $\sqrt2$: This makes the transformation between the two bases unitary, so the eigenvalues of the correlation matrix [and thus formulas like~\eqref{eq: pfaffian reduced dm}] remain unchanged, allowing us to use largely the same code in both cases.

Naturally, entries $H_{ij}$ of the Hamiltonian matrix are expected to be the coefficients of $\gamma_i\gamma_j$, while the correlation matrix is given by $C_{ij}=\unity - \braket{\gamma_i\gamma_j} = \braket{\gamma_j\gamma_i}$.

\paragraph{Specifying the basis in TeMFpy.}

\code{SchmidtModes.from_correlation_matrix()},\\ \code{SchmidtVectors.from_correlation_matrix()}, \code{C_to_MPS()}, \code{C_to_iMPS()}, \code{H_to_MPS()}, and \code{H_to_iMPS()} must know which of the two bases the correlation matrix is specified in. This must be set with the keyword argument \code{basis}, which may be either \code{"M"} (Majorana basis) or \code{"C"} (complex-fermion basis).

\code{correlation_matrix()} can return the correlation matrix in a different basis to the Hamiltonian. This is controlled by the argument \code{basis}, which can be \code{"M->C"} to compute the complex-fermion correlation matrix of a Majorana Hamiltonian, or vice versa with \code{"C->M"}. One can also specify \code{"M->M"} or \code{"C->C"}, which does not change the basis but checks Nambu symmetry in the appropriate basis. (If \code{basis} is not specified, the correlation matrix in the input basis is computed, with no checks.) In addition, helper functions are provided for converting matrices and eigenvectors between the two bases and for checking Nambu symmetry.

Internally, \code{SchmidtModes.from_correlation_matrix()} uses the Majorana basis to diagonalise the correlation matrix, as this makes it easier to ensure Nambu symmetric eigenvectors, see \cref{app: svd pfaffian}. From there on, however, we use the complex-fermion basis, as quantities such as the parity of Bogoliubov vacua or the overlaps~\eqref{eq: pfaffian overlap} are more naturally written in those terms.

\subsubsection{Schmidt decomposition}

\paragraph{Schmidt vectors.}

Given the anomalous correlations $\braket{cc},\braket{c^\dagger c^\dagger}$ in Pfaffian states, we need to diagonalise the full Nambu correlation matrix (rather than just the number-conserving correlators $\langle c^\dagger c\rangle$) restricted to subsystem $A$:
\begin{equation}
    C^{AA} := \begin{pmatrix}
            \langle c_j^\dagger c_i         \rangle & \langle c_j c_i         \rangle \\[2pt]
            \langle c_j^\dagger c_i^\dagger \rangle & \langle c_j c_i^\dagger \rangle
    \end{pmatrix}_{i,j\in A} = N_A\Lambda N_A^\dagger.
\end{equation}
By definition, $C^{AA}$ satisfies the Nambu symmetry $C^{AA} = \unity-\tau^x (C^{AA})^*\tau^x$, whence its eigenvalues are of the form $\lambda_1, \dots, \lambda_n, 1-\lambda_1, \dots,1-\lambda_n$ with $0\le\lambda_\alpha\le 1/2$, while the corresponding eigenvectors can be chosen in the form
\begin{equation}
    N_A = \begin{pmatrix} U & V^*\\V & U^* \end{pmatrix}.
    \label{eq: nambu eigenvectors}
\end{equation}
Now, the reduced density matrix of subsystem $A$ is a mixed fermionic Gaussian mixed state~\cite{Peschel2003GaussianReducedDM,Jin2022BogoliubovMPS}, which can be written in the form
\begin{align}
    \rho_A &= \prod_{\alpha=1}^n \Big[ \lambda_\alpha \gamma_\alpha^\dagger \gamma_\alpha + (1-\lambda_\alpha)\gamma_\alpha \gamma_\alpha^\dagger\Big] &
    (\gamma_\alpha^\dagger, \gamma_\alpha) &= (c_i^\dagger,c_i) \begin{pmatrix} U & V^*\\V & U^* \end{pmatrix}.
    \label{eq: pfaffian reduced dm}
\end{align}
Multiplying this expression out shows that the Schmidt values are products with a term of either $\sqrt{\lambda_\alpha}$ or $\sqrt{1-\lambda_\alpha}$ for each $1\le\alpha\le n$.
The corresponding Schmidt vectors are all Pfaffian states that are annihilated by either $\gamma_\alpha$ (if the corresponding factor in the Schmidt value is $\sqrt{1-\lambda_\alpha}$) or $\gamma_\alpha^\dagger$ (if the corresponding factor in the Schmidt value is $\sqrt{\lambda_\alpha}$).
Since we have defined $\lambda_\alpha\le1/2$, the leading Schmidt state $\ket{\vac_A}$ is annihilated by every $\gamma_\alpha$.
To fix the relative signs of the other Schmidt vectors, we will write them explicitly as $\gamma_\alpha^\dagger\dots\gamma_\beta^\dagger|\vac_A\rangle$.

\paragraph{Consistency relations of left and right Schmidt vectors.}

Similar to the Slater-determinant case, the entangled eigenvectors (that is, those corresponding to $\lambda\neq0$) of $C^{LL}$ and $C^{RR}$ are singular vectors of the off-diagonal block of the Nambu correlation matrix, $C^{LR}$, see \cref{app: svd}. 
However, the singular vectors~\eqref{eq: matching sv} obtained from the SVD of $C^{LR}$ would violate the Nambu symmetry~\eqref{eq: nambu eigenvectors}, which requires a sign difference between $\braket{\gamma^\dagger \gamma}$ and $\braket{\gamma\gamma^\dagger}$ correlators even after diagonalisation. 
Therefore, the closest we can get to an SVD of $C^{LR}$ while respecting Nambu symmetry is
\begin{equation}
    C^{LR} = N_{L,\mathrm{ent}} \begin{pmatrix}
        0 & \diag\left(\sqrt{\lambda_\alpha(1-\lambda_\alpha)}\right) \\
        -\diag\left(\sqrt{\lambda_\alpha(1-\lambda_\alpha)}\right) & 0
    \end{pmatrix} N_{R,\mathrm{ent}}^\dagger,
    \label{eq: pfaffian svd}
\end{equation}
where the subscript ``ent'' indicates the entangled eigenvectors corresponding to $\lambda\neq0$. The ``singular values'' appear in the off-diagonal blocks because the eigenvalues corresponding to a matching pair of left and right singular vectors are $\lambda_\alpha$ and $1-\lambda_\alpha$. This also implies that the nonzero $\lambda_\alpha$ are the same on both sides of the entanglement cut.

In the basis of the $\gamma,\gamma^\dagger$ operators as defined \mbox{in (\ref{eq: pfaffian reduced dm},\, \ref{eq: pfaffian svd})}, the entangled part of the correlation matrix thus consists of $2\times2$ blocks of the form
\begin{align*}
    \begin{pmatrix}
        \braket{\gamma_{L,\alpha}^\dagger\, \gamma_{L,\alpha} } &
        \braket{\gamma_{R,\alpha}\, \gamma_{L,\alpha}}\\[2pt]
        \braket{\gamma_{L,\alpha}^\dagger\, \gamma_{R,\alpha}^\dagger} &
        \braket{\gamma_{R,\alpha} \, \gamma_{R,\alpha}^\dagger}
    \end{pmatrix}
    &=
    \begin{pmatrix}
        \lambda_\alpha & \sqrt{\lambda_\alpha(1-\lambda_\alpha)} \\[2pt]
        \sqrt{\lambda(1-\lambda_\alpha)} & 1-\lambda_\alpha
    \end{pmatrix}\\[1ex]
    \begin{pmatrix}
        \braket{\gamma_{L,\alpha}\, \gamma_{L,\alpha}^\dagger} &
        \braket{\gamma_{R,\alpha}^\dagger\, \gamma_{L,\alpha}^\dagger}\\[2pt]
        \braket{\gamma_{L,\alpha}\, \gamma_{R,\alpha}} &
        \braket{\gamma_{R,\alpha}^\dagger\, \gamma_{R,\alpha}}
    \end{pmatrix}
    &=
    \begin{pmatrix}
        1-\lambda_\alpha & -\sqrt{\lambda_\alpha(1-\lambda_\alpha)}\\[2pt]
        -\sqrt{\lambda(1-\lambda_\alpha)} & \lambda_\alpha
    \end{pmatrix},
\end{align*}
which implies that $\braket{\Gamma_\alpha^\dagger\Gamma_\alpha} = \braket{\tilde\Gamma_\alpha^\dagger \tilde\Gamma_\alpha} = 0$ for
\begin{align*}
    \Gamma_\alpha &= \sqrt{1-\lambda_\alpha}\; \gamma_{L,\alpha} - \sqrt{\lambda_\alpha}\; \gamma_{R,\alpha}^\dagger &
    \tilde\Gamma_\alpha &= \sqrt{\lambda_\alpha}\; \gamma_{L,\alpha}^\dagger + \sqrt{1-\lambda_\alpha}\; \gamma_{R,\alpha},
\end{align*}
that is, that the overall wave function is annihilated by $\Gamma_\alpha$ and $\tilde\Gamma_\alpha$.
This requirement is satisfied by
\begin{equation}
    \ket\psi = \prod_{\lambda_\alpha\neq0} \left(\sqrt{1-\lambda_\alpha} + \sqrt{\lambda_\alpha}\; \gamma_{L,\alpha}^\dagger \gamma_{R,\alpha}^\dagger\right) \ket{\vac_L}\otimes_g \ket{\vac_R},
    \label{eq: pfaffian schmidt decomp}
\end{equation}
which is consistent with~\eqref{eq: pfaffian reduced dm} and fixes all sign ambiguity in the Schmidt decomposition, allowing us to obtain MPS in mixed canonical form.

\subsubsection{Fast overlap computation}
\label{sec: pfaffian fast overlap}

Eq.~\eqref{eq: MPS tensor overlap} gives the MPS tensor entries as overlaps of Pfaffian states; the physical site can be included on the ``bra'' side through an additional Bogoliubov operator $\gamma$, set to either $c_i$ or $c_i^\dagger$ so as to ensure that the vacuum states on the two sides have the same parity (cf.~\Cref{app: pfaffian anticommute}).
Therefore, we need an efficient way of evaluating overlaps of the form
\begin{equation}
    O = \bra{\vac_A} a_{\alpha_k}\dots a_{\alpha_1} b^\dagger_{\beta_1}\dots b^\dagger_{\beta_m}\ket{\vac_B},
    \label{eq: pfaffian overlap def}
\end{equation}
where the $a_\alpha$ and $b_\beta$ are two complete sets of Bogoliubov operators and $\ket{\vac_A}$ and $\ket{\vac_B}$ are the Pfaffian states annihilated by them.
Let the two sets of operators be related to one another by
\begin{equation}
    (b^\dagger,b) = (a^\dagger,a) \begin{pmatrix}
        U & V^* \\ V & U^*
    \end{pmatrix}
    \label{eq: pfaffian formula: operators}
\end{equation}
and assume that $U$ is not singular: this means that $\ket{\vac_{A,B}}$ have the same parity and their overlap is nonzero. Then we have~\cite{Blaizot1985,Robledo2009Onishi}
\begin{equation}
    \ket{\vac_B} = \sqrt{|\det U|}\times \exp\left(\frac12M_{\alpha\beta} a_\alpha^\dagger a_\beta^\dagger\right)\ket{\vac_A},
    \label{eq: pfaffian formula: vacua}
\end{equation}
where $M=(VU^{-1})^*$ is antisymmetric;
the normalisation follows from the Onishi formula~\cite{Robledo1994Onishi}.
We can then show (\Cref{app: pfaffian overlap}) that the overlap~\eqref{eq: pfaffian overlap def} is given by
\begin{subequations}
\label{eq: pfaffian overlap}
\begin{equation}
    O = \sqrt{|\det U|}\times\pf\begin{pmatrix} N^{BB} & N^{BA} \\ -(N^{BA})^\T & N^{AA} \end{pmatrix},
\end{equation}
where
\begin{align}
    N^{AA}_{ij} &= M_{\alpha_i\alpha_j}, &
    N^{BA}_{ij} &= \left[(U^*)^{-1}\right]_{\beta_{m+1-i}\alpha_j}, &
    N^{BB}_{ij} &= [(U^*)^{-1}V]_{\beta_{m+1-i}\beta_{m+1-j}}.
    \label{eq: pfaffian matrix elements}
\end{align}
\end{subequations}
That is, each overlap is given by a Pfaffian the size of which is determined by the number of Bogoliubov excitations in the defining Schmidt vectors, which is capped by the $O(\log\chi)$ number of active generalised orbitals. 
Therefore, computing a full MPS tensor takes $O(N^3+\chi^2\log^3\chi)$ time, just like the Slater-determinant case.
In terms of asymptotic complexity, this matches the approach of Ref.~\cite{Jin2022BogoliubovMPS}; however, instead of a full Bloch--Messiah decomposition, the most complex operations involved here are matrix inversions, which does reduce the CPU time cost of the algorithm.
TeMFpy uses the \code{pfapack} library~\cite{pfapack} to compute the Pfaffians in~\eqref{eq: pfaffian overlap}.

\section{Infinite MPS for gapped systems}
\label{sec: imps}

\begin{figure}[!tp]
    \centering
    \includegraphics{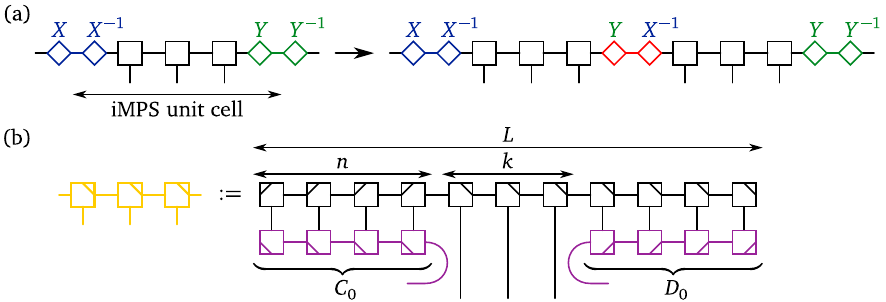}
    \caption{\textbf{(a)} One unit cell's worth of MPS tensors extracted from a finite MPS cannot be used directly as an iMPS unit cell due to the mismatched gauge choices $X$ and $Y$ on its two ends. Upon repeating such a naïve unit cell, these unequal gauge tensors do not cancel out but alter the wave function by inserting spurious unitaries $YX^{-1}$ (red) on the virtual legs connecting unit cells.
    \textbf{(b)} Suitable iMPS unit cells (yellow) can be constructed by contracting the first and last tensors of a unit cell extracted from the longer MPS (black) with unitary matrices $C_0, D_0$ obtained as the overlaps of the left and right environments $\ket{L^{(L,n)}}$ and $\ket{R^{(L, n+k)}}$ with their counterparts $\bra{L^{(L-k,n)}}$ and $\bra{R^{(L-k, n)}}$ in a shorter chain (purple). This transforms the Schmidt bases on both ends of the unit cell to that of the shorter chain at entanglement cut $n$ (free purple legs), thus making them consistent upon repeating. 
    To obtain normalised Schmidt vectors in the environment, both chains must be in mixed canonical form (indicated by diagonal slashes in each tensor block).}
    \label{fig: iMPS}
\end{figure}

\begin{listing}[!t]
\begin{lstlisting}
import numpy as np
from temfpy import slater, iMPS

def H(L, t1=-1, t2=-1.5):
    M = t1 * np.ones(L - 1)
    M[1::2] = t2
    M = np.diag(M, 1)
    return M + M.T

# Control parameters
trunc_par = {"chi_max": 100}  # cf. Listing 1
L_short = 128
cell = 2  # cf. periodicity of H

# Shorter chain
mps_short = slater.H_to_MPS(H(L_short), trunc_par)
# Longer chain
mps_long = slater.H_to_MPS(H(L_short + cell), trunc_par)
# Build the iMPS
imps, error = iMPS.MPS_to_iMPS(
    mps_short, mps_long, sites_per_cell=cell, cut=L_short // 2
)
print(error)
\end{lstlisting}
\vspace{-1ex}
\begin{lstlisting}[language=output]
iMPSError(
    left_unitary=1.04496936e-05,
    left_schmidt=9.77293322e-08,
    right_unitary=1.04564911e-05,
    right_schmidt=5.58724587e-07
)
\end{lstlisting}
    \caption{Basic usage of the \code{temfpy.iMPS} module for a dimerised tight-binding Hamiltonian with a unit cell of 2 sites. The arguments \code{sites_per_cell} ($k$ in \cref{fig: iMPS}) and \code{cut} ($n$) are mandatory and can be provided as positional arguments.}
    \label{listing: iMPS}
\end{listing}

The ground states of gapped, translation-invariant systems can efficiently be represented using matrix product states with finite bond dimensions even in the thermodynamic limit~\cite{Vanderstraeten2019vumps}.
Due to translation invariance, the tensors of such an \textit{infinite} MPS repeat with the periodicity of the Hamiltonian, so the whole infinite system can be captured by MPS tensors for a single unit cell of length $k$.

Naïvely, we could obtain such an iMPS representation by computing the ground state of a sufficiently large finite system and extracting $k$ consecutive tensors $A_{n},\dots,A_{n+k-1}$ near the middle.
However, we have to account for the gauge redundancy of MPS: on any given entanglement cut, redefining the MPS tensors as $A'_i = A_i X$, $A'_{i+1} = X^{-1}A_{i+1}$ leaves the wave function invariant~\cite{Schollwock2011DMRG}.
Even in the limit of very long systems, these gauge choices on entanglement cuts $n$ and $n+k$ will generically be different, so contracting them directly would change the wave function, see~\cref{fig: iMPS}(a).

To construct valid iMPS tensors, we need to fix this gauge freedom.
We follow the approach outlined in Ref.~\cite{Petrica2021MF_MPS}, which effectively replaces the Schmidt vectors $\ket{L^{(L,n)}}$ and $\ket{R^{(L, n+k)}}$, which form the environment of the extracted unit cell, with those of a system one unit cell shorter,  $\ket{L^{(L-k,n)}}$ and $\ket{R^{(L-k, n)}}$:
The latter are defined on the same entanglement cut, so their gauge degrees of freedom are necessarily consistent.
The unitary rotations between these bases are given by 
\begin{align}
    \label{eq: gauge fixing matrices iMPS}
    (C_0)_{\alpha\beta} &= \braket{L_\alpha^{(L-k, n)}|L_\beta^{(L,n)}}, &
    (D_0)_{\alpha\beta} &= \braket{R_\beta^{(L-k,n)}|R_\alpha^{(L,n+k)}}.
\end{align}
As illustrated in \cref{fig: iMPS}(b), replacing the first and last tensors in the unit cell, $A_n$ and $A_{n+k-1}$, with $C_0A_n$ and $A_{n+k-1}D_0$ fixes the gauge on both ends of the unit cell to that of entanglement cut $n$ of the shorter chain.
As such, the unit cell can now be repeated indefinitely without deforming the wave function, i.e., it is a valid iMPS unit cell.
[For finite bond dimensions, the matrices $C_0,D_0$ defined in~\eqref{eq: gauge fixing matrices iMPS} are only approximately unitary: we deal with this issue in \Cref{app: imps}.]

The algorithm outlined above is implemented in the module \code{temfpy.iMPS} for generic MPS, see \Cref{listing: iMPS} for an example workflow.

\subsection{Implementation for mean-field states}
\label{sec: imps mean field}

\begin{listing}[!t]
\begin{lstlisting}
import numpy as np
from temfpy import slater

def H(L, t1=-1, t2=-1.5):
    M = t1 * np.ones(L - 1)
    M[1::2] = t2
    M = np.diag(M, 1)
    return M + M.T

# Control parameters
trunc_par = {"chi_max": 100}  # cf. Listing 1
L_short = 128
cell = 2  # cf. periodicity of H

# Build the iMPS
imps, error = slater.H_to_iMPS(
    H(L_short), H(L_short + cell), trunc_par, sites_per_cell=cell,
    cut=L_short // 2
)
print(error)
\end{lstlisting}
\vspace{-1ex}
\begin{lstlisting}[language=output]
iMPSError(
    left_unitary=8.81564585e-08,
    left_schmidt=9.79503138e-14
)
\end{lstlisting}
    \caption{Usage of the function \code{slater.H_to_iMPS()} for a dimerised tight-binding Hamiltonian with a unit cell of 2 sites. The arguments \code{sites_per_cell} ($k$ in \cref{fig: iMPS}) and \code{cut} ($n$) are mandatory and can be provided as positional arguments. The calling sequence of \code{pfaffian.H_to_iMPS()} is identical, except that one must specify the full Nambu Hamiltonian/correlation matrix in the input (cf.\ \cref{sec: pfaffian basis}).}
    \label{listing: slater iMPS}
\end{listing}

For Slater determinants and Pfaffian states, however, it is not necessary to construct the Schmidt states surrounding the iMPS unit cell as MPS, as their overlaps~\eqref{eq: gauge fixing matrices iMPS} can efficiently be computed the same way as MPS tensor entries, cf.~\cref{sec: slater fast overlap,sec: pfaffian fast overlap}.
The classes \code{slater.MPSTensorData} and \code{pfaffian.MPSTensorData} are able to compute such overlaps too.

\code{slater.C_to_iMPS()} and \code{pfaffian.C_to_iMPS()} thus compute the MPS tensors of the longer chain for a single unit cell of the iMPS \textit{only}, as well as the left and right Schmidt vectors for the reference entanglement cut $n$ of the shorter chain.
The unitary rotation $C$ on the left end of the iMPS unit cell is applied the same way as explained above and in \Cref{app: imps} for a generic MPS.
By contrast, as we compute right canonical iMPS tensors, the unitary rotation at the right end can be combined into the construction of the last tensor by using the Schmidt vectors of the shorter chain~\cite{Petrica2021MF_MPS}:
\begin{align}
    \tilde B^{n_k}_{\alpha\beta} &= B^{n_k}_{\alpha\gamma} (D_0)_{\gamma\beta} = 
    \braket{R_\beta^{(L-k,n)}|R_\gamma^{(L,n+k)}}\times \left(\langle R^{(L,n+k)}_\gamma| \otimes_g \langle n_k|\right)\,
    |R^{(L,n+k-1)}_\alpha \rangle \nonumber \\
    &= \left(\langle R^{(L-k,n)}_\beta| \otimes_g \langle n_k|\right)\,
    |R^{(L,n+k-1)}_\alpha \rangle,
\end{align}
(summation over Greek Schmidt-vector indices implied) because the Schmidt vectors $\ket{R^{(L,n+k)}}$ form a complete orthonormal basis. 
As such, these routines do not return any estimate of the conversion error introduced on the right end of the iMPS unit cell.

An example workflow with these routines is shown in \Cref{listing: slater iMPS}. 
As only a fraction of all MPS tensors needs to be computed, this approach is much faster than that in \Cref{listing: iMPS}.
Furthermore, it avoids the truncation errors introduced by converting the environments of the iMPS unit cell to MPS form (cf.~the error estimates in \Cref{listing: iMPS,listing: slater iMPS}).

\section{Gutzwiller projection}
\label{sec: gutzwiller}

In addition to studying them directly or using them as starting points for DMRG on fermionic systems, mean-field wave functions are also useful building blocks for studying such exotic systems as quantum spin liquids or fractional quantum Hall states~\cite{Becca2017QuantumSystems,Wen2002QuantumLiquids,Wen2007QuantumElectrons,Savary2017QuantumReview,Jain1989Parton,Lu2012PartonFCI,McGreevy2012PartonFCI,Li2025PartonSUnk}.
These are often described by parton constructions, where the physical degrees of freedom are split up into a number of fractionalised quasiparticles, which capture the physics of the strongly entangled phase already at mean-field level.
This construction, however, enlarges the Hilbert space on each lattice site: 
To recover a wave function of the original system, the additional degrees of freedom must be removed by acting with a \textit{Gutzwiller projection} operator on every site.
This is quite a natural operation for matrix product states, as the Gutzwiller projector can be applied directly to the physical leg(s) associated with each site, resulting in a (generally no longer canonical) MPS with unchanged virtual space.

\begin{listing}[!t]
\begin{lstlisting}
import numpy as np
from temfpy import slater, gutzwiller

# Hamiltonian, in this case a 10-site tight-binding chain
H = -(np.eye(10, k=1) + np.eye(10, k=-1))

# Ground-state correlation matrix of one spin species
C, _ = slater.correlation_matrix(H)

# Unprojected MPS without particle-hole rotation
mps_fermion = slater.C_to_MPS(C, {"chi_max": 100}, spinful="simple")
# Projection without particle-hole rotation
mps_spin = gutzwiller.abrikosov(mps_fermion)
print("MPS site type without particle-hole rotation:", mps_spin.sites[0])

# With particle-hole rotation
mps_fermion = slater.C_to_MPS(C, {"chi_max": 200}, spinful="PH")
mps_spin = gutzwiller.abrikosov_ph(mps_fermion)
print("MPS site type with particle-hole rotation:", mps_spin.sites[0])
\end{lstlisting}
\vspace{-1ex}
\begin{lstlisting}[language=output]
MPS site type without particle-hole rotation: SpinHalfSite('None')
MPS site type with particle-hole rotation: SpinHalfSite('Sz')
\end{lstlisting}
\caption{Usage of the \code{temfpy.gutzwiller} module for spinon mean-field states of $U(1)$ quantum spin liquids. Starting from the correlation matrix of the Slater determinant of up-spin spinons \code{C}, projected states can be built with or without the partial particle-hole transformation~\eqref{eq: PH rotation}. In the latter case, the resulting spin MPS preserves the $S^z$ quantum number. For $\mathbb{Z}_2$ spin liquids, the mean-field state can still be written as a Slater determinant in the particle-hole rotated convention, but the construction from a single spin species using the \code{spinful} switch is not applicable.}
\label{listing: gutzwiller}
\end{listing}

There are many Gutzwiller projection schemes, appropriate for different parton constructions. 
As an illustration of the general principle, we implemented Gutzwiller projection for quantum spin liquids in spin-$1/2$ systems, which are often described using mean-field theories of Abrikosov fermions:
\begin{align}
    \vec S_i &= \frac12 f^\dagger_{i\alpha} \vec{\sigma}\!_{\alpha\beta} f_{i\beta}, &
    (f^\dagger_{i\alpha}f_{i\alpha} &= 1)
\end{align}
where $\vec\sigma$ are the Pauli matrices and Greek indices (summation implied) stand for the Abrikosov-fermion species $\uparrow,\downarrow$. 
The on-site Hilbert space of the Abrikosov fermions is four-dimensional, which is reduced to the spin-$1/2$ Hilbert space by imposing single occupancy on each site.
We implement this projection scheme in \code{temfpy.gutzwiller.abrikosov()}. 
The input MPS is expected to have $2L$ spinless fermion sites:
sites $2i$ and $2i+1$ represent modes $f_{i\uparrow}$ and $f_{i\downarrow}$, respectively.
The function groups these pairs of sites, projects out the unphysical (empty/doubly occupied) sector of each on-site Hilbert space, and reinterprets the remaining two-dimensional Hilbert space as a spin-$1/2$ degree of freedom.
Since the only conserved quantum number of the input MPS is (the parity of the) fermion number, which is fixed by the projection, the output spin MPS will not carry any symmetry quantum numbers.

Spin liquids in Heisenberg models, however, preserve spin-rotation symmetry, which implies that the Abrikosov-fermion mean-field theories that describe them may only contain singlet hopping and pairing terms~\cite{Wen2002QuantumLiquids,Baskaran1987TheTheory,Baskaran1988GaugeSystems}.
Similar to singlet superconductivity~\cite{Becca2017QuantumSystems}, these terms can all be treated as hopping if we perform a particle--hole transformation on one spin species:
\begin{align}
    c_{i,\uparrow} &:= f_{i,\uparrow}, &
    c_{i,\downarrow} &:= f_{i,\downarrow}^\dagger.
    \label{eq: PH rotation}
\end{align}
In terms of the $c$ operators, the valid spin Hilbert space corresponds to empty (spin down) and fully occupied (spin up) sites.%
\footnote{We also note that the full gauge freedom of spin liquids is most naturally expressed in terms of mixing the $c$ particles~\cite{Affleck1988SU2Model,Wen2002QuantumLiquids}.}
This Gutzwiller projection is implemented in\\ \code{temfpy.gutzwiller.abrikosov_ph()}, whose input sites are expected to correspond to the fermionic modes $c_{i\uparrow}$ and $c_{i\downarrow}$.
Since in this convention, the fermion occupation number corresponds to magnetisation along the $\sigma^z$ axis, number-conserving (Slater-determinant) states are projected to $U(1)$ spin-preserving MPS.

$U(1)$ quantum spin liquids can be represented using mean-field Hamiltonians containing only spin-independent hopping terms, $H_\mathrm{MF}=\sum_{ij\sigma} t_{ij} f_{i\sigma}^\dagger f_{j\sigma}$~\cite{Wen2002QuantumLiquids,Wen2007QuantumElectrons,Savary2017QuantumReview}.
The correlation matrices of such states consist of decoupled spin-up and spin-down blocks that are either equal (in the $f$ convention) or related by $C_\uparrow=\unity-C_\downarrow$ (in the $c$ convention).
For convenience, we provide the function \code{temfpy.slater.spinful_correlation_matrix()} that turns the correlation matrix for a single spin species into such an enlarged correlation matrix, compatible with either convention.
This function can also be applied to the input correlation matrix by passing the argument \code{spinful="simple"} ($f$ convention) or \code{spinful="PH"} ($c$ convention) to \code{temfpy.slater.C_to_MPS()} etc.

An illustration of this functionality is given in \Cref{listing: gutzwiller}.
We have not tried to build a catalogue of all useful Gutzwiller projection schemes with this first release.
However, we welcome and look forward to users who deploy TeMFpy to other strongly correlated systems enriching the library by contributing their Gutzwiller projection schemes.

\section[Example workflow: kagome Dirac spin liquid]{Example workflow: Gutzwiller-projected wave functions and flux threading for the kagome Heisenberg model}
\begin{figure}[!tp]
    \centering
    \includegraphics{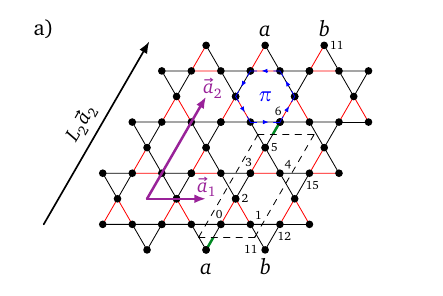}%
    \includegraphics{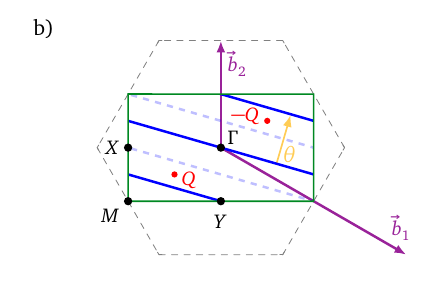}
    \caption{\textbf{(a)} The $\pi$-flux kagome tight-binding model: Black (red) bonds correspond to $\varphi_{ij} = 0$ ($\pi$) in~\eqref{eq: pi flux ham}.
    We consider twisted boundary conditions, where winding around the cylinder once also moves one unit cell forward along the $x$ direction (cf.\ identified sites $a$ and $b$ on the top and bottom).
    As a result, sites $i$ and $i+6$ have identical environments, so iMPS unit cells of size 6 can be used for a cylinder of arbitrary width.
    \\
    \textbf{(b)} Brillouin zones of the standard kagome lattice (gray) and of the $\pi$-flux model (green). The Dirac points are located at $\pm Q$. Dark blue lines indicate allowed values of the wave vector for the geometry shown in (a).
    Adding a spin flux $\theta$ along the cylinder translates these lines perpendicularly to themselves (orange arrow).
    Dashed light-blue lines indicate the allowed momenta for $\theta = 2\pi$, which give rise to an alternative candidate spin-liquid state.}
    \label{fig: pi-flux-model-geometry}
\end{figure}

The most promising candidate for the ground state of the antiferromagnetic spin-1/2 Heisenberg model 
\begin{equation}
    \hat{H} = J \sum_{\langle ij \rangle} \hat{\mathbf{S}}_i \cdot \hat{\mathbf{S}}_j
    \label{eq: kagome Hamiltonian}
\end{equation}
on the kagome lattice is a gapless $U(1)$ Dirac spin liquid (DSL)~\cite{hasting2000DiracKagome, Wen2007KagomeHeisenberg, Yasir2011KagomeHeisenberg, pollman2017SignaturesOfDirac}.
As an illustrative example for a typical workflow with TeMFpy, we consider the following hopping-only parton ansatz for the DSL:
\begin{equation}
    \label{eq: pi flux ham}
    \hat{H}_{\pi-\mathrm{flux}} = - \sum_{\sigma=\uparrow,\downarrow} \sum_{<ij>} e^{i\varphi_{ij}} f^\dagger_{i, \sigma} f_{j, \sigma} \equiv \sum_{\alpha \beta} f^\dagger_\alpha h_{\alpha \beta} f_\beta,
\end{equation}
where the $f_{i,\sigma}$ are a pair of Abrikosov fermions on site $i$, Greek indices stand for the combination of site and spin species, and the phases $\varphi_{ij}$ are chosen such that the total flux for a parton traversing a hexagon (triangle) is $\pi$ (0). 
Implementing this flux pattern requires extending the unit cell to six sites; we use the phase pattern shown in \cref{fig: pi-flux-model-geometry}(a).
The resulting band structure has two Dirac cones per spin species [\cref{fig: pi-flux-model-geometry}(b)].

\subsection{Spin-flux threading}

Ref.~\cite{pollman2017SignaturesOfDirac} identified Dirac cones in the spin model~\eqref{eq: kagome Hamiltonian} using spin-flux threading, that is, by inserting a spin rotation by angle $\theta$ into the periodic boundary conditions around the cylinder:
\begin{align}
    S^\pm_{x,y,s} &= e^{\pm i\theta}S^\pm_{x+\Delta,y+L_2,s}; &
    S^z_{x,y,s} &= S^z_{x+\Delta,y+L_2,s},
    \label{eq: spin twist full ring}
\end{align}
where sites are indexed with the position $x,y$ of the containing 6-site unit cell in units of the lattice vectors $\vec a_{1,2}$ [\cref{fig: pi-flux-model-geometry}(a)], and the index $0\le s<6$ of the site within the unit cell, and $L_2$ is the number of 6-site unit cells around the cylinder.
The periodicity of the cylinder does not have to line up with the lattice direction $\vec a_2$;
rather, winding around the cylinder may move the particle along the $\vec a_1$ direction by $\Delta$ unit cells.
In particular, we will use $L_2=2$ and $\Delta=1$~\cite{pollman2017SignaturesOfDirac}: 
The latter means that consecutive six-site unit cells form a spiral on which each unit cell is surrounded by identical environments, that is, we can use an iMPS unit cell of 6 sites regardless of $L_2$.
Furthermore, we can split the spin rotation~\eqref{eq: spin twist full ring} uniformly between the $L_2$ unit cells by deforming the Heisenberg Hamiltonian as
\begin{equation}
    \hat H(\theta) = J\sum_{\langle ij \rangle} \left(\hat S_i^z \hat S_j^z + \frac12 \left[\hat S_i^+ \hat S_j^- e^{\theta[\mathrm{cell}(i)-\mathrm{cell}(j)]/L_2} + \mathrm{h.c.}\right] \right),
    \label{eq: kagome spin twisted hamiltonian}
\end{equation}
where $\mathrm{cell}(i)$ stands for the index of the unit cell containing site $i$ in the spiral.
Writing~\eqref{eq: kagome spin twisted hamiltonian} in terms of Abrikosov fermions and performing the same mean-field decoupling that generates~\eqref{eq: pi flux ham} in the untwisted case yields
\begin{equation}
    \hat{H}_{\pi-\mathrm{flux}}(\theta) = - \sum_\sigma \sum_{<ij>} e^{i\varphi_{ij}} f^\dagger_{i, \sigma} f_{j, \sigma} e^{\sigma\theta[\mathrm{cell}(i)-\mathrm{cell}(j)]/2L_2},
    \label{eq: kagome spin twisted parton}
\end{equation}
where $\sigma=\uparrow,\downarrow$ stand for $+1,-1$ in the exponential~\cite{pollman2017SignaturesOfDirac}.
In other words, the effective Hamiltonians of up- (down-)spin partons acquire flux $\theta/2$ ($-\theta/2$) for winding around the cylinder.

The eigenstates of a single-particle Hamiltonian on the cylinder are the same as those on the open plane, provided the wave function is compatible with the periodic boundary conditions. 
This requirement selects lines in $k$-space parallel to the length of the cylinder; adding a phase twist to the boundary conditions shifts the lines of allowed momenta around the cylinder [\cref{fig: pi-flux-model-geometry}(b)].
In particular, for our geometry, the lines of allowed momenta pass through the Dirac cones of the $\pi$-flux Hamiltonian~\eqref{eq: pi flux ham} for $\theta=\pi$.
At this point, correlations become algebraic even on the finite-width cylinder; due to the Dirac dispersion of excitations, the correlation length should diverge as $\xi^{-1}\sim|\theta-\pi|$. 
This inverse correlation length can be extracted from an iMPS transfer matrix as the logarithm of its leading eigenvalue, $\xi^{-1} = \log \lambda$~\cite{Schollwock2011DMRG, pollman2017SignaturesOfDirac}. 

\subsection{The TeMFpy workflow}

\paragraph{Building the mean-field Hamiltonian.}

The TeMFpy workflow starts by writing out the quadratic fermion Hamiltonian $h_{\alpha\beta}$ as an explicit matrix.
For an iMPS representation in particular, we need this Hamiltonian for two cylinders whose lengths differ by the desired iMPS unit cell and which are long enough for the iMPS unit cell extracted from the middle of the cylinder to be representative of the thermodynamic limit.
Since the Hamiltonian is block diagonal in spin space and $h_\downarrow(\theta)=\bigl(h_\uparrow(\theta)\bigr)^*$, we only need to construct $h_\uparrow(\theta)$:%
\footnote{For the sake of readability, we omit the construction of the Hamiltonian here. The unabridged example can be found \href{https://temfpy.github.io/temfpy/tutorials/pi_flux_kagome.html}{in the online documentation.}}
\begin{lstlisting}
import numpy as np

def generate_sp_hamiltonian(L1, L2, spinflux):
    """Constructs the single-particle mean-field Hamiltonian of up-spin
    spinons for a cylindrical spiral of L1 x L2 unit cells with Delta = 1."""

L1, L2, spinflux = 50, 2, np.pi/2

H_long_spinless = generate_sp_hamiltonian(L1, L2, spinflux)
# all unit cells are identical, so simply cut the last one off
H_short_spinless = H_long_spinless[:-6, :-6]
\end{lstlisting}

\paragraph{Computing the correlation matrices.}

We start this step by computing the correlation matrix of $h_\uparrow(\theta)$ at half-filling using \code{temfpy.slater.correlation_matrix()}.
Next, we need to expand this to a correlation matrix for partons of both spin; as explained in Sec.~\ref{sec: gutzwiller}, we also perform a particle-hole rotation on the down spins so that the $U(1)$ particle number correspond to $S^z$.
Usually, this would be handled by \code{temfpy.slater.spinful_correlation_matrix()}; however, in our case, $h_\uparrow(\theta)$ and $h_\downarrow(\theta)$ are not equal due to the spin flux insertion, so we implement
\begin{equation}
    \label{eq: pi flux spinfulC}
    \left[C^{(\mathrm{spinful)}}_{\mathrm{ph}}(\theta)\right]_{ij} =
    \begin{pmatrix}
        C_\uparrow(\theta)_{ij} & 0 \\
        0 &  \delta_{ij} - C_\downarrow(\theta)_{ij}
    \end{pmatrix} =
    \begin{pmatrix}
        C_\uparrow(\theta)_{ij} & 0 \\
        0 &  \delta_{ij} - C_\uparrow(\theta)^*_{ij}
    \end{pmatrix}
\end{equation}
manually:
\begin{lstlisting}
from temfpy.slater import correlation_matrix

def spinful_C_with_conjugation(H):
    L = len(H)
    assert L%2 == 0 # otherwise half-filling is impossible

    C_up, _= correlation_matrix(H, L//2)
    # Add down-spin part
    C = np.zeros((2*L, 2*L), dtype=C_up.dtype)
    C[::2, ::2] = C_up
    C[1::2, 1::2] = np.eye(L) - C_up.conj()
    return C

C_short_spinful = spinful_C_with_conjugation(H_short_spinless)
C_long_spinful = spinful_C_with_conjugation(H_long_spinless)
\end{lstlisting}
Note that the up- and down-spin Abrikosov fermions on each site must be next to one another in the MPS to make Gutzwiller projection possible.

\paragraph{Mean-field and Gutzwiller-projected iMPS.}

We now reach the main functionality of TeMFpy, the conversion of mean-field states to MPS form. 
As explained in \cref{sec: imps}, we have to specify which unit cell of the longer mean-field state should be extracted as a representative of the thermodynamic limit. 
This unit cell should align with the repeating unit cell of the mean-field Hamiltonian (in our case, it should start and end at integer multiples of the $2\times6=12$-site unit cell) and be as close to the centre of the longer cylinder as possible:
\begin{lstlisting}
from temfpy.slater import C_to_iMPS
from temfpy.gutzwiller import abrikosov_ph

cut = 12 * ((L1 * L2 - 1) // 2) # Start of middle unit cell in C_short
psi_parton, err = C_to_iMPS(
    C_short = C_short_spinful, 
    C_long = C_long_spinful, 
    trunc_par = {'chi_max': 2000}, 
    sites_per_cell = 12, # size of iMPS unit cell to extract
    cut = cut,
)

# Gutzwiller projection
# Creates new MPS object by default; consume input MPS with inplace=True
# Converts to canonical form by default; disable with return_canonical=False 
psi_spin = abrikosov_ph(psi_parton, inplace=False, return_canonical=True)
\end{lstlisting}
A high bond dimension should be used for constructing the mean-field state, since the full fermionic state contains more degrees of freedom and is hence more entangled than the spin state left behind after Gutzwiller projection.

\paragraph{Correlation-length spectra.}

\begin{figure}[!tp]
    \centering
    \includegraphics{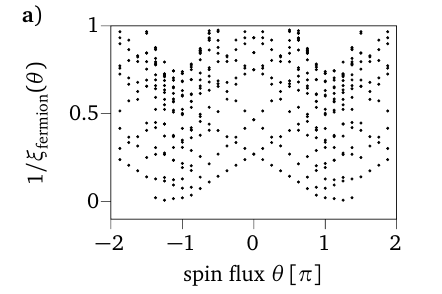}%
    \includegraphics{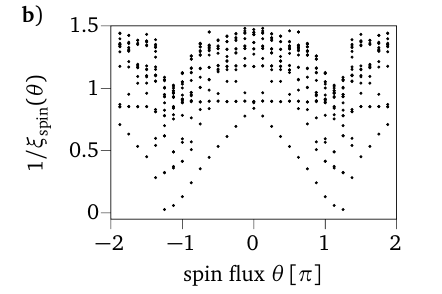}
    \includegraphics{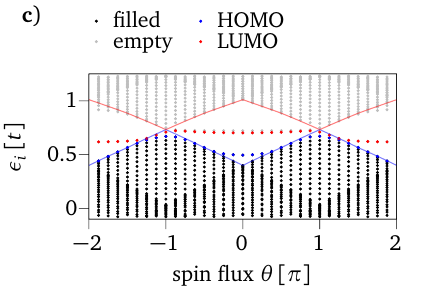}%
    \includegraphics{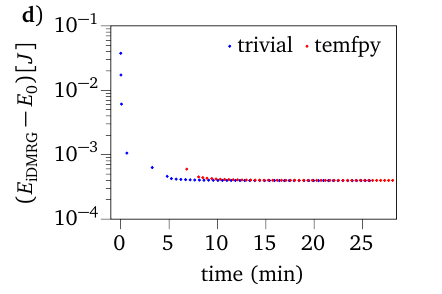}
    \caption{
    \textbf{(a,\,b)}
    TeMFpy-generated iMPS transfer-matrix spectra of the mean-field state of up-spin partons [\textbf{(a)}; bond dimension: 300] and of the Gutzwiller-projected Dirac-spin-liquid state [\textbf{(b)}; bond dimension: 2000].
    Compared to iDMRG simulations~\cite{pollman2017SignaturesOfDirac}, the Dirac points extend into a wide, nearly gapless region at $|\theta| \gtrsim \pi$.\\
    \textbf{(c)}
    Single-particle energy levels of the up-spin partons in~\eqref{eq: kagome spin twisted parton} for a finite cylinder.
    The highest occupied (HOMO) and lowest unoccupied (LUMO) levels at half-filling are highlighted in blue and red, respectively. Band edges of an infinite cylinder are indicated by the solid lines.\\
    \textbf{(d)}
    Evolution of the iDMRG ground-state energy as a function of wallclock time starting from the Gutzwiller-projected DSL state obtained using TeMFpy (red; initial bond dimension before projection: 2000) and the $S^z$ product state $\ket{\uparrow \downarrow \uparrow \downarrow \uparrow \downarrow}$ (blue).
    We used TeNPy's two-site iDMRG implementation with bond dimension 1000 on a cluster node with 6 cores and 16 GB memory.
    The reference energy $E_0=-0.4376$ was taken from~\cite{pollman2017SignaturesOfDirac} (bond dimension $6000$). \\
    All simulations use the geometry shown in \cref{fig: pi-flux-model-geometry}; TeMFpy used cylinders of length $L_1=50$ unit cells to build the iMPS unit cells.
    The code used to generate these figures is available in the \href{https://github.com/temfpy/projects/tree/main/pi_flux_kagome}{TeMFpy projects repository.}
    }
    \label{fig: pi_flux_results}
\end{figure}

As explained above, we now look for Dirac-cone signatures in the $\theta$-dependence of the correlation lengths associated with the parton-parton correlator $\langle c_{i\uparrow}^\dagger c_{j\uparrow}\rangle$ in the mean-field Hamiltonian $H_\uparrow$ [\cref{fig: pi_flux_results}(a)] as well as the spin-spin correlator $\langle S_i^+ S_j^-\rangle$ in the Gutzwiller-projected DSL state [\cref{fig: pi_flux_results}(b)].
Since these operators change the conserved quantum numbers of the iMPS (particle number and $2S^z$, respectively), the relevant eigenvalues of the iMPS transfer matrix are those associated with eigenvectors of the appropriate nonzero charge (1 and 2, respectively).
For example, the spin-spin correlation lengths are obtained as follows:
\begin{lstlisting}
from tenpy.networks import TransferMatrix

tm = TransferMatrix(psi_spin, psi_spin, charge_sector=[2])
num_ev = 20
eigV, _ = tm.eigenvectors(num_ev = num_ev)
inverse_correlation_length = -np.log(np.abs(eigV))
\end{lstlisting}
The overall structure of both spectra agrees with the iDMRG results of~\cite{pollman2017SignaturesOfDirac}.
However, a qualitative difference appears near $\theta = \pi$: 
The spectrum of correlation lengths is asymmetrical around this point, with inverse correlation lengths going well below those found with DMRG for $|\theta|>\pi$.
We attribute this to the different geometries underlying the iMPS construction in the two methods:
While iDMRG works directly in the thermodynamic limit, TeMFpy extracts the iMPS unit cell from the bulk of a finite cylinder, which has (non-topological) edge modes in its single-particle spectrum. These give rise to a wide, nearly gapless region at $|\theta| \gtrsim \pi$, see Fig.~\ref{fig: pi_flux_results}(c).

\paragraph{Gutzwiller-boosted iDMRG.}

The iMPS generated with TeMFpy can also serve as an initial state for iDMRG.
Previous studies~\cite{Jin2021GutzwillerDMRG} have shown that mean-field initial states can improve convergence and help avoid local minima in finite-size DMRG.
For iDMRG on the kagome Heisenberg model, however, the $S^z$ product state $\psi_\mathrm{trivial} = \ket{\uparrow \downarrow \uparrow \downarrow \uparrow \downarrow}$ and the Gutzwiller-projected $\pi$-flux ansatz converge to the same ground-state energy [\cref{fig: pi_flux_results}(d)].
The latter starts from a significantly lower initial energy, confirming the parton model~\eqref{eq: pi flux ham} as a good candidate description of the true ground state;
however, the cost of computing the initial iMPS environment for a state with a high bond dimension (note the very long runtime of the first iDMRG iteration) outweighs this benefit for iDMRG.
We note, however, that this behaviour is model-dependent: In particular, it is much harder to write down good fermionic trial states, making the ability to convert Hartree--Fock states to MPS form particularly useful for strongly-correlated-electron systems.

\section{Conclusion}

We have presented TeMFpy, a Python library for converting fermionic mean-field states to MPS form.
The library incorporates functionality for both Slater determinants and Pfaffian states, as well as finite and infinite MPS representations.
We have aimed to make the key functionality readily accessible to users without understanding the details of the algorithm.
Nevertheless, the library is fully modular, allowing users to easily prototype new applications of our algorithms.
While the algorithms we use are not entirely new~\cite{Petrica2021MF_MPS,Jin2022BogoliubovMPS,Liu2025EfficientGaussianMPS,Li2025EfficientGaussianMPS},
we have devised and incorporated several improvements that make the calculations both more efficient and numerically stable.
Equally importantly, an easy-to-use open-source implementation of these algorithms, based around an existing tensor-network library~\cite{tenpy}, opens up its potential to a wide range of potential users.

In the future, we will improve and expand the capabilities of TeMFpy, while maintaining a clean, user-friendly, and backwards-compatible interface. 
A particularly interesting prospect is incorporating the approach of~\cite{Liu2025EfficientGaussianMPS} adapted to bosonic Gaussian states, where our linear-algebra manipulations fail, but by introducing virtual boson degrees of freedom, the problem might still be reduced to computing permanents of a manageable size.
Furthermore, we will expand the Gutzwiller-projection facility in \code{temfpy.gutzwiller} to a wider range of parton constructions. In particular, we plan to design a general interface that allows users to prototype the projection schemes relevant to their work without a detailed understanding of TeNPy.
In addition to TeNPy, we would like to support generating MPS in formats suitable for other tensor-network libraries, such as ITensor~\cite{itensor}.
Last but not least, we will also expand the worked examples published with the codebase in order to illustrate more workflows and to help users get started with similar projects.

In all of this, we are looking forward to the support of the community of users, which we hope will build up around TeMFpy, as it will be deployed for studying (Gutzwiller-projected) mean-field states, seeding DMRG/VUMPS, and other applications we do not yet anticipate.

\section*{Acknowledgements}
We thank Hao Chen, Johanna Ockenfels, and Hong-Hao Tu for helpful discussions and Lucas Zeifang for comprehensively testing the library and bringing many bugs to our attention.

\paragraph{Funding information}
S.\,H.\,H. and A.\,Sz. were supported by Ambizione grant No. 215979 by the Swiss National Science Foundation.

\begin{appendix}
\numberwithin{equation}{section}

\section{Finding the highest Schmidt values}
\label{app: highest schmidt values}

The Schmidt values implied by the Schmidt decompositions \eqref{eq: generalSlaterDet} and \eqref{eq: pfaffian schmidt decomp} are both of the form
\begin{equation}
    \lambda(P) = \prod_{i \in P} a_i \prod_{j \in \overline{P}} b_j
\end{equation}
for some $a_1,\dots,a_n$ and $b_1,\dots,b_n$,
where $n$ is the number of entangled orbitals, $P$ is a subset of $\{1,2,\dots,n\}$, and $\overline P = \{1,2,\dots,n\}\backslash P$.
Therefore, finding the most significant $\chi$ Schmidt vectors is equivalent to finding the subsets $P_1,\dots,P_\chi$ for which $\lambda(P)$ is the highest.
Defining $M_i = \max(a_i,b_i)$ and $m_i =\min(a_i,b_i)$, $\lambda(P)$ for any subset $P$ can also be written as
\begin{align}
    \lambda(P) = \tilde\lambda(P') &= \prod_{i \in P'} m_i \prod_{j \in \overline{P}'} M_j = \prod_{i\in P'}\frac{m_i}{M_i} \underbrace{\prod_{j=1}^n M_j}_{\lambda_\mathrm{max}}\\
    \log\tilde\lambda(P') &= \log\lambda_\mathrm{max} - \sum_{i\in P'} \underbrace{\log\left(\frac{M_i}{m_i}\right)}_{r_i}
\end{align}
for another subset $P'\subseteq\{1,\dots,n\}$. That is, finding the highest Schmidt values is equivalent to finding the subsets of $\{r_1,\dots,r_n\}$ with the lowest sum. Note that all $r_i\ge 0$ by construction.

\begin{algorithm}[tb]
\begin{algorithmic}[1] 
    \Statex \textbf{Input:}
    $[r_1,\dots,r_n]$, an array of nonnegative numbers, sorted in increasing order. 
    \Function{lowestSum}{$[r_1,\dots,r_n]$}
    \State $S = [\{\}]$ \Comment{The very lowest sum must be handled specially}
    \State initialize min-heap $H$ \Comment{Ordered by subset sum}
    \State H.push$( \{ 1\} ,r_{1} ,1)$ \Comment{Second lowest sum}
    \For{$i=2,\dots, 2^n$}
        \State $(P_i,t_i,\ell)$ $\gets$ $H$.min 
        \State $H$.pop$(P_i,t_i,\ell)$
        \State append $P_i$ to $S$
        \If{$\ell < n$}
            \State $H$.push$(P_i \cup \{\ell+1\}, t_i + r_{\ell+1}, \ell+1)$ \Comment{Descendants contain $\ell$}
            \State $H$.push$(P_i \cup \{\ell+1\} \backslash \, \{\ell\},t_i + r_{\ell+1} - r_{\ell},  \ell+1)$\Comment{Descendants don't contain $\ell$}
        \EndIf
    \EndFor
    \State \Return $S$
\EndFunction
\end{algorithmic}
\caption{Subsets of a set of nonnegative numbers in order of increasing sum. Note that the main loop can be terminated at any $\chi<2^n$ if only the $\chi$ lowest sums are required.}
\label{algo: lowest sum}
\end{algorithm}

In the following we show that \Cref{algo: lowest sum}, given in~\cite{LowestSums} without proof, indeed generates every subset in order of increasing sum. 
We first note that every tuple $(P,t,\ell)$ pushed on the heap satisfies $t = \sum_{i\in P} r_i$ and $\ell = \max (P)$; it is easy to see that lines 4, 10 and 11 obey these relations.

\paragraph{All subsets are obtained.}

We can view the subsets $P_i \cup \{\ell+1\}$  and $P_i \cup \{\ell+1\} \backslash \, \{\ell\}$ pushed to the heap on lines 10 and 11 as children of $P_i$ that are located on level $(\ell+1)$ of an $n$-level binary tree.
That is, sets with highest entry $K$ can only appear on level $K$ of this binary tree.
To see that every nonempty\footnote{The empty subset corresponds to the lowest possible sum. We deal with it separately on line 2.} subset of $\{1,\dots,n\}$ does appear in the tree, we observe that $\{k\}$ is always added to the subset at level $k$ and it may only be removed at level $k+1$.
Therefore, subset $P$ can only appear on level $K=\max(P)$ of the binary tree, on the node obtained by taking line 10 at levels $\{k+1: k\in P, k < K\}$ and line 11 on all other levels.
As such, the $2^n-1$ nonempty subsets of $\{1,\dots,n\}$ are in one-to-one correspondence with the nodes of the binary tree, i.e., every subset is generated precisely once.

\paragraph{Subsets are returned in order of increasing sum.}

Since every subset of $\{1,\dots,n\}$ is returned at some point, this is equivalent to showing that the sum $t_i$ returned in each step is no less than the sum $t_{i-1}$ in the previous step. 
$0=t_1\le t_2=r_1$ holds by definition.
For $i> 2$, we have two cases to consider:
    If $P_i$ was already contained in the heap when $P_{i-1}$ was popped, then $t_{i-1} \leq t_{i}$ by the heap property.
    Otherwise, it is one of the children of $P_{i-1}$ in the binary tree introduced above, pushed on the heap in step $i-1$. By assumption, $r_{\ell}\geq 0$ and $r_{\ell} - r_{\ell-1} \geq 0$, so both children of $P_{i-1}$ have a sum no less than $t_{i-1}$. Therefore, $t_i\ge t_{i-1}$ again.

\paragraph{Complexity.}

If only the $\chi$ subsets with the lowest sums are required, the inner loop of \Cref{algo: lowest sum} can clearly be terminated at step $i=\chi$ instead of running to $2^n$. 
The memory required by the algorithm up to this point is determined by the total number of entries in the return array $S$ and the heap $H$, which is in turn the total number of entries pushed to the heap (they are either still on the heap, or have been moved to $S$). In each iteration of the loop, at most two entries can be pushed, therefore, the total number of pushes is $\Theta(\chi)$. Since each entry contains a subset of $\{1,\dots,n\}$, their total size, i.e., the memory complexity is $\Theta(\chi n)$.

The time complexity of the algorithm is dominated by the push and pop operations of the heap. By the arguments above, the heap size at step $i$ is $\Theta(i)$, so in a basic binary heap each of these operations take $\Theta(\log i)$ time. Integrating this from $i=2$ to $\chi$ gives time complexity $\Theta(\chi\log \chi)$.

\section{Mutual diagonalisation of $C^{LL}$, $C^{LR}$, and $C^{RR}$}
\label{app: svd}

The correlation matrices of both Slater determinants and Pfaffian states must satisfy $C^2=C$~\cite{Blaizot1985}:
Intuitively, all eigenvalues of $C$ must be either 1 (corresponding to filled orbitals or Bogoliubov operators that annihilate the Pfaffian) or 0 (corresponding to empty orbitals or Bogoliubov excitations). 
This also implies that the eigenvalues of the diagonal submatrices of $C$ are between~0 and~1.

Upon introducing an entanglement cut that splits the system into a left ($L$) and a right ($R$) half, it is natural to split the correlation matrix into blocks:
\begin{equation}
    C = \begin{pmatrix}
        C^{LL} & C^{LR} \\ (C^{LR})^\dagger &C^{RR}
    \end{pmatrix}
\end{equation}
(note that $C$ is Hermitian by definition),
for which $C^2=C$ implies
\begin{subequations}
\label{eq: mutual diag}
\begin{align}
    (C^{LL})^2 + C^{LR}(C^{LR})^\dagger &= C^{LL}
    \label{eq: CLL}\\
     C^{LL} C^{LR}+ C^{LR} C^{RR} &= C^{LR}
    \label{eq: CRL}\\
    (C^{LR})^\dagger C^{LR} + (C^{RR})^2 &= C^{RR};
    \label{eq: CRR}
\end{align}
\end{subequations}
as $C$ is Hermitian, the $(C^{LR})^\dagger$ block does not yield an independent constraint.

Let $u$ be an eigenvector of $C^{LL}$ with eigenvalue $\lambda$. Then by~\eqref{eq: CLL}, $u$ is also an eigenvector of $C^{LR}(C^{LR})^\dagger$ with eigenvalue $\lambda(1-\lambda)$, i.e., it is a left singular vector of $C^{LR}$ with singular value $\sigma = \sqrt{\lambda(1-\lambda)}$. Likewise, \eqref{eq: CRR} implies that every eigenvector of $C^{RR}$ is a right singular vector of $C^{LR}$.

Let now $0<\lambda<1$, i.e., let $u$ define an entangled orbital. Then the normalised right singular vector corresponding to $u$ is
\begin{align}
    v &= \frac{(C^{LR})^\dagger u}\sigma &
    u &= \frac{C^{LR} v}\sigma.
    \label{eq: matching sv}
\end{align}
Now, multiplying~\eqref{eq: CRL}$^\dagger$ from the right by $u/\sigma$ yields
\begin{align}
    \frac{(C^{LR})^\dagger C^{LL} u}\sigma + C^{RR} v &= v \nonumber\\
    C^{RR} v &= (1-\lambda)v.
    \label{eq: lambda 1-lambda}
\end{align}
That is, the eigenvalues of $C^{LL}$ and $C^{RR}$ corresponding to matching left and right singular vectors of $C^{LR}$ sum to 1.

Finally, noting that the eigenbases of $C^{LL}$ and $C^{RR}$ provide a full complement of left and right singular vectors of $C^{LR}$ and that $\sigma\neq0$ only for the entangled modes $0<\lambda<1$, we have the following singular-value decomposition of $C^{LR}$:
\begin{align}
    C^{LR} = \sum_\mathrm{ent} \sqrt{\lambda(1-\lambda)}\; u_\lambda\, v_{1-\lambda}^\dagger,
\end{align}
where the subscripts indicate the eigenvalues corresponding to each eigenvector, and $u$ and $v$ are related by~\eqref{eq: matching sv}.

\subsection{Implementation for Slater determinants}
\label{app: svd slater}
In case both left and right Schmidt vectors are needed for a single entanglement cut, the phase relation~\eqref{eq: matching sv} between the entangled eigenvectors of $C_{LL}$ and $C_{RR}$ must be fixed in addition to diagonalising them.

Suppose that the columns of the matrix $\tilde{U}$ contain the eigenvectors of $C^{LL}$ with eigenvalue $\lambda$, while $\tilde{V}$ contains the eigenvectors of $C^{RR}$ with eigenvalue $1 - \lambda$.
Then~\eqref{eq: mutual diag} implies
\begin{equation}
   \tilde{C}^{LR} := \tilde{U}^\dagger C^{LR} \tilde{V} = \sigma (1 - \sigma) W,
\end{equation}
where $W$ is an arbitrary unitary matrix.
We fix this unitary degree of freedom by an additional SVD within the degenerate block:
\begin{align}
    \tilde{C}^{LR} &= P\Sigma Q^\dagger \implies &
    U&:= \tilde U P &
    V&:= \tilde V Q.
    \label{eq: svd gauge fix}
\end{align}
$U$ and $V$ are still eigenvectors of $C^{LL}$ and $C^{RR}$ with respective eigenvalues $\lambda$ and $1-\lambda$, but in addition, they are also matching singular vectors of $C^{LR}$.
Doing this for all degenerate blocks, we obtain all corresponding pairs of entangled Schmidt modes. To make the best use of vectorisation in NumPy, we group the degenerate subspaces by multiplicity, so that all matrices $U,V,\tilde C^{LR},\dots$ of the same size may be handled together.

\subsection{Implementation for Pfaffian states}
\label{app: svd pfaffian}

Degenerate eigenspaces of $C^{LL}$ and $C^{RR}$ are handled largely the same way for Pfaffian states. Maintaining the Nambu symmetry~\eqref{eq: nambu eigenvectors}, however, requires two modifications:
\begin{itemize}
    \item Generically, the numerical diagonalisation of $C^{LL}$ and $C^{RR}$ does not return eigenvectors that respect~\eqref{eq: nambu eigenvectors}.
    Therefore, we perform the gauge fixing~\eqref{eq: svd gauge fix} only for eigenspaces $\tilde U:=N_{L,\mathrm{ent},\lambda}$ and $\tilde V:=N_{R,\mathrm{ent},1-\lambda}$ with $\lambda<1/2$.
    Then, we enforce~\eqref{eq: nambu eigenvectors} by replacing $N_{L,\lambda>1/2}$ and $N_{R,\lambda<1/2}$ with the Nambu counterparts of the transformed $N_{L,\lambda<1/2}$ and $N_{R,\lambda>1/2}$.
    (This latter is also done for the non-entangled modes.)
    \item For $\lambda=1/2$, the Nambu-partner eigenvectors are degenerate.
    In this case, it is convenient to work in the Majorana basis, where these pairs of eigenvectors are simply complex conjugates of one another: 
    This implies that their real and imaginary parts are also eigenvectors and orthogonal to one another. 
    
    That is, the $\lambda=1/2$ eigenspace (of dimension $d$) has a complete real basis.
    Numerically, we obtain this by an SVD of the concatenated real and imaginary parts of the generically complex eigenvectors:
    To numerical accuracy, $d$ of the singular values is 1, corresponding to the real eigenvectors $r_\alpha$; the remaining $d$ singular values are 0.

    In the Majorana basis, $C^{LR}$ is purely imaginary, so $r_L^\dagger \Im(C^{LR}) r_R$ has a purely real SVD, which can be used to transform $r_L,r_R$ so that $r_L^\dagger C^{LR} r_R = i/2\times \unity_d$. From these, finally, orthonormal and Nambu-symmetric eigenvectors of $C^{LL}$ and $C^{LR}$ satisfying~\eqref{eq: pfaffian svd} can be obtained as
    \begin{align*}
        n_{L,\alpha} &= \frac{r_{L,\alpha}+ir_{L,\alpha+d/2}}{\sqrt2}, &
        \overline{n_{L,\alpha}} &= \frac{r_{L,\alpha}-ir_{L,\alpha+d/2}}{\sqrt2},\\
        n_{R,\alpha} &= \frac{ir_{R,\alpha}+r_{R,\alpha+d/2}}{\sqrt2}, &
        \overline{n_{R,\alpha}} &= \frac{-ir_{R,\alpha}+r_{R,\alpha+d/2}}{\sqrt2}.
        & (0\le \alpha < d/2)
    \end{align*}
\end{itemize}

\section{Handling fermion anticommutation}

\subsection{Slater determinants}

\paragraph{Expanding~\eqref{eq: generalSlaterDet}.}

\begin{figure}[t]
    \centering
    \includegraphics{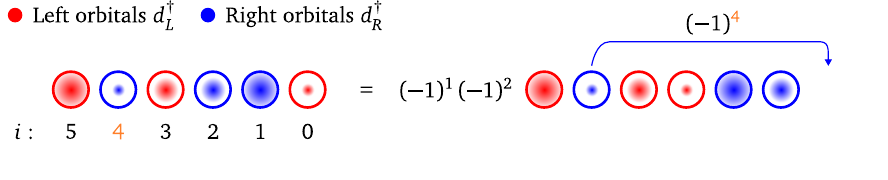}
    \caption{Separating left and right entangled operators $d_L^\dagger, d_R^\dagger$ in one term of the Schmidt decomposition~\eqref{eq: generalSlaterDet}. 
    Starting point is the entangled orbitals sorted in decreasing order of $\lambda$ (the magnitude of $\lambda$ or $(1-\lambda)$ is indicated by the sizes of the inner red or blue blobs, respectively).
    The right entangled operators are then swapped one by one to the right end, starting from the rightmost.
    The resulting fermionic anticommutation sign for each right entangled orbital is given by $(-1)^i$, where $i$ denotes the initial position in the list of entangled orbitals, counted from the right.}
    \label{fig: comm slater exp}
\end{figure}

For a proper Schmidt decomposition we need to multiply~\eqref{eq: generalSlaterDet} out, taking care of the fermionic anticommutation signs.
A typical term takes the form
\begin{align}
    \ket{S} &= \prod_{\mu \in O_L} d^\dagger_{L,\mu} \left(
    \dots d^\dagger_{R,i+1} d^\dagger_{L,i} d^\dagger_{R,i-1} \dots \right) \prod_{\nu \in O_R} d^\dagger_{R,\nu} \ket{\mathrm{vac}}  \\
    &= 
    \mathrm{sgn(\pi)} \left[\prod_{\mu \in O_L} d^\dagger_{L,\mu} \left(\dots d^\dagger_{L,i} \dots\right) \ket{\mathrm{vac}_L} \right]
    \otimes_g 
    \left[\left(\dots d^\dagger_{R,i-1} d^\dagger_{R,i+1} \dots\right) \prod_{\nu \in O_R} d^\dagger_{R,\nu} \ket{\mathrm{vac}_R} \right] ,
\end{align}
where $\pi$ is the permutation that permutes all entangled operators $d^\dagger_R$ through the entangled operators $d^\dagger_L$ and reverses their order.

One can build $\pi$ by moving the right entangled orbitals one by one to the right.
Due to their reversed order, the number of swaps needed for each orbital depends only on its initial position $i$ in the list of entangled orbitals, where the indexing starts from 0 on the right, see \cref{fig: comm slater exp}.
To account for the resulting anticommutation signs, we multiply each entangled orbital operator $d^\dagger_{R,i}$ by $(-1)^i$.

\paragraph{Reclassification of entangled orbitals.}

\begin{figure}[t]
    \centering
    \includegraphics{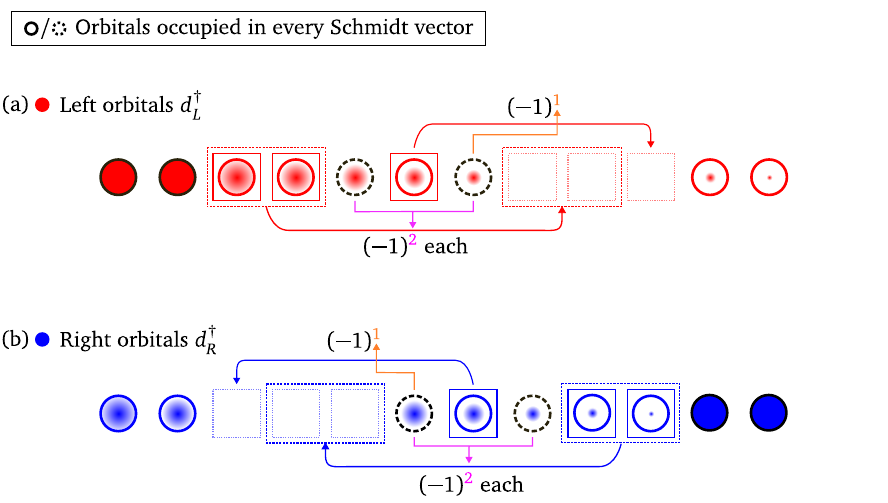}
    \caption{To include entangled orbitals that are in fact occupied in every Schmidt vector (dashed black border and partially filled) in the submatrix $A$ in~\eqref{eq: block matrix determinant formula}, all entangled operators $d_L^\dagger$ \textbf{(a)} / $d_R^\dagger$ \textbf{(b)} with mixed occupation (red/blue border) have to be moved past them.
    Starting from the entangled orbital with the smallest/largest eigenvalue (indicated by the size of the inner blob), they are moved one by one to the right/left, without changing their relative order. 
    Therefore, each entangled orbital picks up a fermionic anticommutation sign determined by the number of reclassified orbitals to its right/left.}
    \label{fig: comm slater reclass}
\end{figure}

If the bond dimension is limited to $\chi$, some entangled orbitals may in fact be occupied in every Schmidt vector we keep.
To be able to include them in the submatrix $A$ in~\eqref{eq: block matrix determinant formula}, these need to be commuted through the remaining entangled orbitals, introducing an anticommutation sign.
It is useful to view this as commuting all remaining entangled orbitals one by one through the newly classified ones.
The anticommutation sign incurred in each of these steps depends only on the number of occupied orbital operators it needs to be commuted through, see \cref{fig: comm slater reclass}:
We account for these signs by multiplying them to the corresponding orbital eigenvectors.

\subsection{Pfaffian states}
\label{app: pfaffian anticommute}

\paragraph{Expanding~\eqref{eq: pfaffian schmidt decomp}.}
For a proper Schmidt decomposition, we need to multiply~\eqref{eq: pfaffian schmidt decomp} out, taking care of the fermionic anticommutation signs. 
A typical term takes the form
\begin{align}
    \ket{S} = \gamma_{L,\alpha_1}^\dagger\gamma_{R,\alpha_1}^\dagger\dots \gamma_{L,\alpha_k}^\dagger\gamma_{R,\alpha_k}^\dagger \ket{\vac_L}\otimes_g \ket{\vac_R} 
    &= \gamma_{L,\alpha_1}^\dagger\dots \gamma_{L,\alpha_k}^\dagger\gamma_{R,\alpha_k}^\dagger\dots\gamma_{R,\alpha_1}^\dagger \ket{\vac_L}\otimes_g \ket{\vac_R},
\end{align}
which differs from~\eqref{eq: schmidt decomp general} in that the operators and states of the left- and right-half Hilbert spaces are not separated from one another.
That is, we need to commute the $k$ $\gamma_R^\dagger$ operators across $\ket{\vac_L}$. If the latter is parity odd, each of these steps incur a negative sign~\cite{Bultinck2017FermionicMPS}, so we end up with
\begin{align}
    \ket S = (-1)^{k \pi_L} \left(\gamma_{L,\alpha_1}^\dagger\dots \gamma_{L,\alpha_k}^\dagger\ket{\vac_L} \right) \otimes_g \left(\gamma_{R,\alpha_k}^\dagger\dots\gamma_{R,\alpha_1}\ket{\vac_R}\right),
\end{align}
where $\pi_L=0,1$ is the parity of $\ket{\vac_L}$, computed using the Bloch--Messiah decomposition~\cite{BlochMessiah1962}.
To account for this additional fermionic sign in the TeMFpy implementation, we
\begin{itemize}
    \item list the $\gamma^\dagger$ operators in the definition of left and right Schmidt states in opposite order (namely, in the code, we list them in increasing order of $\lambda_\alpha$ on the left and in decreasing order on the right);
    \item and flip the sign of all $\gamma_R,\gamma_R^\dagger$ operators if the Bogoliubov vacuum $\ket{\vac_L}$ is parity odd.
\end{itemize}

\paragraph{MPS tensor entries as Pfaffian overlaps.}

The overlaps to evaluate for~\eqref{eq: MPS tensor overlap} are not exactly of the form~\eqref{eq: pfaffian overlap def}, but rather
\begin{subequations}
\begin{align}    
     A^{n_i}_{\alpha\beta} &= \bra{0_{i}} c_i^{n_i} \otimes_g \bra{\vac_L^{(i-1)}} \gamma_{L,\alpha_k}^{(i-1)}\dots \gamma_{L,\alpha_1}^{(i-1)} \otimes_g \gamma_{L,\beta_1}^{(i)\dagger}\dots \gamma_{L,\beta_m}^{(i)\dagger}\ket{\vac_L^{(i)}}\\
     B^{n_i}_{\beta\alpha} &= \bra{\vac_R^{(i)}} \gamma_{R,\alpha_k}^{(i)}\dots \gamma_{R,\alpha_1}^{(i)}  \otimes_g \bra{0_{i}} c_i^{n_i} \otimes_g \gamma_{R,\beta_1}^{(i-1)\dagger}\dots \gamma_{R,\beta_m}^{(i-1)\dagger}\ket{\vac_R^{(i-1)}}.
\end{align}
\end{subequations}
As before, we need to commute $\bra{\vac_L^{(i-1)}}$ or $\bra{0_{i}}$ through a number of Bogoliubov operators:
\begin{subequations}
\begin{align}    
     A^{n_i}_{\alpha\beta} &= (-1)^{n_i\pi_L^{(i-1)}}\bra{0_{i}} \otimes_g \bra{\vac_L^{(i-1)}} c_i^{n_i} \gamma_{L,\alpha_k}^{(i-1)}\dots \gamma_{L,\alpha_1}^{(i-1)} \gamma_{L,\beta_1}^{(i)\dagger}\dots \gamma_{L,\beta_m}^{(i)\dagger}\ket{\vac_L^{(i)}}
     \label{eq: pfaffian mps commute A}\\
     B^{n_i}_{\beta\alpha} &= \bra{\vac_R^{(i)}} \otimes_g \bra{0_{i}} \gamma_{R,\alpha_k}^{(i)}\dots \gamma_{R,\alpha_1}^{(i)} c_i^{n_i} \gamma_{R,\beta_1}^{(i-1)\dagger}\dots \gamma_{R,\beta_m}^{(i-1)\dagger}\ket{\vac_R^{(i-1)}},
     \label{eq: pfaffian mps commute B}
\end{align}
\end{subequations}
where $\pi_L^{(i-1)}$ is the parity of $\ket{\vac_L^{(i-1)}}$; Eq.~\eqref{eq: pfaffian mps commute B} incurs no extra sign as $\ket{0_i}$ is parity even.
To account for the additional sign in~\eqref{eq: pfaffian mps commute A}, we use $-c_i$ rather than $c_i$ as additional Bogoliubov annihilation operator if $\ket{\vac_L^{(i-1)}}$ is parity odd.

\paragraph{Parity matching for Pfaffian overlaps.}

The Pfaffian overlap formula~\eqref{eq: pfaffian overlap} is only valid if the vacua $\ket{\vac_A}$ and $\ket{\vac_B}$ have the same fermion parity, which is not necessarily the case for Schmidt states on different entanglement cuts.
We remedy this problem by particle-hole transforming one Bogoliubov operator $a_0$ on the bra side: $a_0 \leftrightarrow a_0^\dagger$, $\bra{\vac_A'}:=\bra{\vac_A}a_0$.
For numerical stability, we choose the most entangled (i.e., $\lambda$ closest to $1/2$, or the on-site mode for MPS tensor entries) Bogoliubov mode: 
This defines $\ket{\vac_A'}$ as the highest-weight bra Schmidt state of the same parity as the leading ket Schmidt state, which are expected to have a high overlap in our use cases.
Furthermore, across all Schmidt vectors, this mode is expected to be excited the most often, so flipping it leads to the smallest possible increase in the size of the Pfaffians that need to be computed.

For left Schmidt vectors, the Bogoliubov mode operators are listed in order of increasing $\lambda$, so the most entangled mode $a_0$ appears closest to $\bra{\vac_A}$ in the string of operators~\eqref{eq: pfaffian overlap def}. 
Upon the particle-hole transformation, bra Schmidt states with or without an $a_0$ Bogoliubov excitation transform as
\begin{subequations}
\begin{align}
    \bra{\vac_A} a_0 a_{\alpha_k}\dots a_{\alpha_1} &=\bra{\vac_A'} a_{\alpha_k}\dots a_{\alpha_1}\\
    \bra{\vac_A} a_{\alpha_k}\dots a_{\alpha_1} &= \bra{\vac_A} a_0a_0^\dagger a_{\alpha_k}\dots a_{\alpha_1} = \bra{\vac_A'} a'_0 a_{\alpha_k}\dots a_{\alpha_1},
\end{align}
\end{subequations}
without generating any additional fermionic sign.

For right Schmidt vectors, the Bogoliubov mode operators are listed in order of decreasing $\lambda$, so the most entangled mode $a_0$ appears at the end of the string of $a_\alpha$ in~\eqref{eq: pfaffian overlap def}. 
Upon the particle-hole transformation, bra Schmidt states with or without an $a_0$ Bogoliubov excitation transform as
\begin{subequations}
\begin{align}
    \bra{\vac_A}  a_{\alpha_k}\dots a_{\alpha_1} a_0 &= (-1)^k \bra{\vac_A}  a_0 a_{\alpha_k}\dots a_{\alpha_1} = (-1)^k \bra{\vac_A'} a_{\alpha_k}\dots a_{\alpha_1}\\
    \bra{\vac_A} a_{\alpha_k}\dots a_{\alpha_1} &= \bra{\vac_A}  a_{\alpha_k}\dots a_{\alpha_1}a_0a_0^\dagger = (-1)^k \bra{\vac_A'} a_{\alpha_k}\dots a_{\alpha_1}a'_0 .
\end{align}
\end{subequations}
To account for this additional fermionic sign, we flip the sign of every bra Schmidt mode operator other than $a_0$.

\section{Proof of the Pfaffian overlap formula}
\label{app: pfaffian overlap}

In the following, we use the Bogoliubov operators $a_\alpha,b_\beta$ and vacua $|\vac_{A,B}\rangle$ defined in \mbox{Eqs. (\ref{eq: pfaffian formula: operators},\,\ref{eq: pfaffian formula: vacua})}. We start with a simpler result than~\eqref{eq: pfaffian overlap}:

\paragraph{Lemma 1.}
Let $\gamma_i$ ($1\le i\le n$) be either $a_\alpha$ or $a_\alpha^\dagger$ for some $\alpha$. Then
\begin{equation}
    O := \frac{ \bra{\vac_A} \gamma_n\dots\gamma_2\gamma_1\ket{\vac_B}}{\sqrt{|\det U|}} = \pf N,
    \label{eq: bog overlap basic}
\end{equation}
where the antisymmetric $n\times n$ matrix $N$ is defined by
\begin{equation}
    N_{i>j} = \begin{cases}
        0 & \text{$\gamma_i = a^\dagger_\alpha$ for some $\alpha$}\\
        -\delta_{\alpha_i\alpha_j} &\gamma_i=a_{\alpha_i}, \gamma_j = a_{\alpha_j}^\dagger\\
        M_{\alpha_i\alpha_j}&\gamma_i=a_{\alpha_i}, \gamma_j = a_{\alpha_j}.
    \end{cases}
    \label{eq: overlap pfaffian elements basic}
\end{equation}

\begin{proof}
By induction on the number of $a^\dagger$ operators among the $\gamma_i$.
If all $\gamma_i = a_{\alpha_i}$ for all $i$, 
\begin{equation*}
    O = \bra{\vac_A} a_{\alpha_n}\dots a_{\alpha_1} \exp\left(\frac12M_{\alpha\beta} a_\alpha^\dagger a_\beta^\dagger\right)\ket{\vac_A} = \pf\, [M_{\alpha_i\alpha_j}]_{ij}
\end{equation*}
is well known. Now, let there be $k>0$ $a^\dagger$ among the $\gamma_i$ and let the leftmost of these be $\gamma_j = a_{\alpha_j}^\dagger$.
Let $\ket{S} = |\det U|\times\gamma_{j-1}\dots \gamma_1\ket{\vac_B}$ for brevity. 
Let us commute $a_{\alpha_j}^\dagger$ through to the left:
\begin{align*}
    O &= \bra{\vac_A} a_{\alpha_n}\dots a_{\alpha_{j+1}} a_{\alpha_j}^\dagger \ket S\\
    &= \delta_{\alpha_{j+1}\alpha_j} \bra{\vac_A} a_{\alpha_n}\dots a_{\alpha_{j+2}} \ket S - \bra{\vac_A} a_{\alpha_n}\dots a_{\alpha_{j+2}} a_{\alpha_j}^\dagger a_{\alpha_{j+1}} \ket S = \dots\\
    &= \sum_{i=j+1}^n (-1)^{i+j+1} \delta_{\alpha_i\alpha_j}  \bra{\vac_A} a_{\alpha_n}\dots a_{\alpha_{i+1}}a_{\alpha_{i-1}}\dots a_{\alpha_{j+1}} \ket S,
\end{align*}
since $\bra{\vac_A}a_{\alpha_j}^\dagger = 0$. The expectation values on the last line contain $(k-1)$ $a^\dagger$ operators, so by the induction step, they are given by the Pfaffian of $[N_{\backslash ij}]$, i.e., the matrix $N$ with columns and rows $i$ and $j$ removed. By definition, $\delta_{\alpha_i\alpha_j} = N_{ji}$ for $j<i$ ($N$ is antisymmetric), and $N_{ji}=0$ for $j\ge i$.  Therefore,
\begin{align*}
    O = \sum_{i=j+1}^n (-1)^{i+j+1} N_{ji} \pf\,[N_{\backslash ij}],
\end{align*}
which is precisely the expansion of $\pf N$ by its $j$th row/column. 
\end{proof}

In the case of interest, the operators in $O$ are not simply $a$ or $a^\dagger$, but rather their linear combinations. However, the result is still a single Pfaffian:

\paragraph{Lemma 2.} Let $\gamma_i= a_\alpha C_{\alpha i} + a_\alpha^\dagger D_{\alpha i}$ for $1\le i\le n$. Then, 
\begin{align}
    O &:= \bra{\vac_A} \gamma_n\dots\gamma_2\gamma_1\ket{\vac_B} = \sqrt{|\det U|}\times{\pf N},
    \label{eq: bog overlap general}
\end{align}
where the antisymmetric $n\times n$ matrix $N$ is defined by
\begin{equation}
    N_{i>j} = C_{\cdot i}^\T  (MC_{\cdot j} - D_{\cdot j}).
    \label{eq: overlap pfaffian elements general}
\end{equation}

\begin{proof}
We consider the intermediate case where $\gamma_i$ for all $i>k$ are either $a_{\alpha_i}$ or $a_{\alpha_i}^\dagger$, i.e., there is precisely one nonzero element of 1 in $C_{\cdot i}$ and $D_{\cdot i}$ together, and proceed by induction on $k$. 
$k=0$ is precisely Lemma 1.

For $0<k\le L$, we use that the overlap~\eqref{eq: bog overlap general} is linear in $\gamma_k$, so we have
\begin{align*}
    O &= \sum_\alpha C_{\alpha k} \bra{\vac_A} \gamma_n\dots\gamma_{k+1} a_{\alpha}\gamma_{k-1}\dots \gamma_1\ket{\vac_B} + D_{\alpha k} \bra{\vac_A} \gamma_n\dots\gamma_{k+1} a_{\alpha}^\dagger\gamma_{k-1}\dots \gamma_1\ket{\vac_B}.
\end{align*}
By the induction step, these are all Pfaffians that only differ in their $k$th row/column.
Since the Pfaffian too is linear in its rows/columns, $O$ is also a Pfaffian, whose $k$th row/column is the linear combination of these, with coefficients $C_{\cdot k}, D_{\cdot k}$. The claim follows from the fact that~\eqref{eq: overlap pfaffian elements general} is also linear in $C_{\cdot k}, D_{\cdot k}$.
\end{proof}

Eq.~\eqref{eq: pfaffian overlap} is a special case of Lemma~2 with
\begin{equation*}
    \gamma_i = \begin{cases}
        b_{\beta_{m+1-i}}^\dagger = a_\alpha U_{\alpha\beta_{m+1-i}} + a_\alpha^\dagger V_{\alpha\beta_{m+1-i}} & 1\le i\le m\\
        a_{\alpha_{i-m}} & m < i \le m+k,
    \end{cases}
\end{equation*}
for which the overlap can be written as 
\begin{equation}
    \bra{\vac_A} a_{\alpha_k}\dots a_{\alpha_1} b^\dagger_{\beta_1}\dots b^\dagger_{\beta_m}\ket{\vac_B} = \sqrt{|\det U|} \times  \pf\begin{pmatrix} N^{BB} & N^{BA} \\ N^{AB} & N^{AA} \end{pmatrix},
\end{equation}
where
\begin{subequations}
\label{eq: overlap pfaffian elements}
\begin{align}
    N^{AA}_{ij} &= M_{\alpha_i\alpha_j}
    \label{eq: NAA}\\
    N^{AB}_{ij} &= (M V_{\cdot\beta_{m+1-j}} - U_{\cdot\beta_{m+1-j}})_{\alpha_i}
    \label{eq: NAB}\\
    N^{BB}_{ij} &= V_{\cdot\beta_{m+1-i}}^\T  (M V_{\cdot\beta_{m+1-j}} - U_{\cdot\beta_{m+1-j}})
    \label{eq: NBB}
\end{align}
\end{subequations}
and $N^{BA} = -(N^{AB})^\T $.
Note that~\eqref{eq: overlap pfaffian elements general} applies as written to all entries of $N^{AB}$ as all $b^\dagger$ operators precede all $a$ operators.
We can also directly obtain the lower triangle $i>j$ of $N^{BB}$, but~\eqref{eq: NBB} turns out to be antisymmetric: $V^\T MV$ is antisymmetric because $M$ is, while $V^\T U+U^\T V=0$ follows from the unitarity of the Bogoliubov transformation matrix.
We can simplify~\eqref{eq: NAB} and~\eqref{eq: NBB} further by noting that $M$ is antisymmetric:
\begin{align*}
    M &= (VU^{-1})^* = -(VU^{-1})^\dagger = -(U^\dagger)^{-1}V^\dagger\\
    MV-U &= -(U^\dagger)^{-1}V^\dagger V -U= -(U^\dagger)^{-1}\left(\unity - U^\dagger U\right)-U = \left[U-(U^\dagger)^{-1}\right]-U = -(U^\dagger)^{-1},
\end{align*}
so we have
\begin{subequations}
\begin{align}
    N^{AB}_{ij} &= -\left[(U^\dagger)^{-1}\right]_{\alpha_i\beta_{m+1-j}}\\
    N^{BB}_{ij} &= -\left[V^\T (U^\dagger)^{-1}\right]_{\beta_{m+1-i}\beta_{m+1-j}} = [(U^*)^{-1}V]_{\beta_{m+1-i}\beta_{m+1-j}};
\end{align}
\end{subequations}
the last equality follows because $N^{BB}$ is antisymmetric.

\section{Practical implementation of the iMPS conversion}
\label{app: imps}

\begin{figure}[t]
    \centering
    \includegraphics{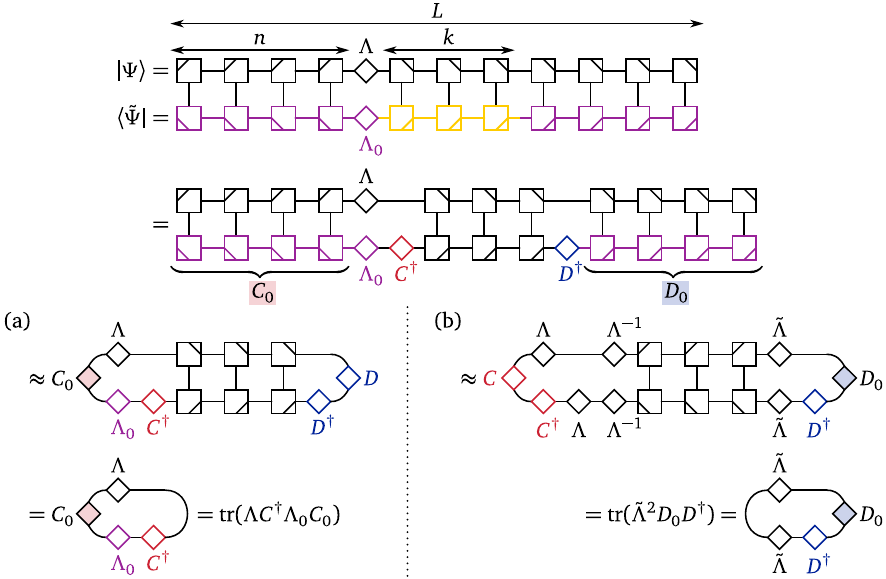}
    \caption{Overlap of the wave function $\ket\Psi$ on the longer chain of length $L$ and its reconstruction $\ket{\tilde\Psi}$ from MPS tensors of a shorter chain of length $L-k$ and a single iMPS unit cell of length $k$.
    Diagonal slashes indicate whether each tensor is left- or right-canonical.
    \textbf{(a,\,b)} Estimates of the overlap assuming that the reconstruction of the environment to the right \textbf{(a)} or to the left \textbf{(b)} of the iMPS unit cell is perfect.}
    \label{fig: imps approximations}
\end{figure}

For finite bond dimensions, the Schmidt vectors $\ket{L_\alpha^{(L,n)}}$ and $\ket{L_\alpha^{(L-k,n)}}$ (or $\ket{R_\beta^{(L, n+k)}}$ and $\ket{R_\beta^{(L-k, n)}}$) do not span the whole Hilbert space, so, generically, the overlap matrices~\eqref{eq: gauge fixing matrices iMPS} are not unitary.
This means that instead of rotating the tensors at the ends of the iMPS unit cell using $C_0, D_0$ directly, we have to find appropriate unitary rotations $C$ and $D$.

Furthermore, for finite systems, the Schmidt values $\Lambda$ and $\Lambda_0$ are not exactly equal for two different finite system sizes, adding an additional source of mismatch between the wave function $\ket\Psi$ on the longer chain, and its reconstruction $\ket{\tilde\Psi}$ using the wave function on the shorter chain and the iMPS unit cell. In the following, we will find the unitary matrices $C$ and $D$ that maximise the overlap $\braket{\tilde\Psi|\Psi}$ (\cref{fig: imps approximations}).

For the sake of concreteness, we assume from here on that the iMPS unit cell is right canonical (this choice is built into TeMFpy as well).
We will also assume that errors introduced by the iMPS construction on the left and right ends of the unit cell are independent, that is, when looking for the optimal $C$, we assume that the optimal unitary $D$ is $D_0$ and vice versa.

To find the optimal left gauge rotation matrix $C$, we can thus contract down all the right canonical tensors in the right environment and the iMPS unit cell [\cref{fig: imps approximations}(a)]. That is, the overlap of the two wave functions is given by $\tr( \Lambda_0 C_0\Lambda C^\dagger)$, which can be maximised using the following result:

\paragraph{Lemma.}
Let the singular value decomposition of a square matrix $A$ be $A=U\Sigma V^\dagger$ and let all singular values be strictly positive. Then, the unitary matrix $R$ for which $\tr(AR^\dagger)$ is maximal is $R=UV^\dagger$.

\begin{proof}
    Let $R$ be a unitary matrix. Then $\tr(AR^\dagger) = \tr(U\Sigma V^\dagger R^\dagger) = \tr(\Sigma V^\dagger R^\dagger U) = \sum_i \sigma_i (V^\dagger R^\dagger U)_{ii}$.
    Since $V^\dagger R^\dagger U$ is unitary, its diagonal elements can be at most 1. 
    Therefore, $\tr(AR^\dagger)$ is highest if all diagonal elements of $V^\dagger R^\dagger U$ are 1, i.e., if it is the identity matrix. Rearranging yields $R=UV^\dagger$.
\end{proof}

Finally, the smallest possible deviation between $\ket\Psi$ and $\ket{\tilde\Psi}$ is
\begin{align}
    \big\Vert \ket\Psi - \ket{\tilde\Psi}\big\Vert^2 
    &= \tr\Lambda^2+\tr\Lambda_0^2 - \tr( C_0\Lambda C^\dagger \Lambda_0) -\tr(\Lambda_0 C\Lambda C_0^\dagger)
    \nonumber\\
    &= \tr\left[(\Lambda_0 C - C_0\Lambda)\times\mathrm{H.c.}\right] + \tr\left[\Lambda (\unity-C_0^\dagger C_0)\Lambda \right],
    \label{eq: iMPS error left}
\end{align}
where we used that $C$ is unitary. 
Both terms are nonnegative:
the first is the Frobenius norm of $\Lambda_0 C - C_0\Lambda$;
as for the second, diagonal entries of $C_0^\dagger C_0$ are at most 1 (equal to 1 if $C_0$ is unitary).
In fact, we may assign distinct meanings to the two error terms:
\begin{itemize}
    \item The second term measures the deviation of the overlap matrix $C_0$ from being unitary, weighted with the corresponding Schmidt values. This error term can be reduced by keeping more Schmidt vectors, i.e., using higher bond dimensions.
    \item The first term, by contrast, measures the deviation of the Schmidt values $\Lambda$ and $\Lambda_0$ and how accurately $C$ maps between them. This error term can be reduced by using larger system sizes, where finite-size differences between $\ket{L_\alpha^{(L,n)}}$ and $\ket{L_\alpha^{(L-k,n)}}$ are smaller.
\end{itemize}

Likewise, to find the optimal right gauge rotation matrix $D$, we neglect the error introduced on the left hand side; in particular, we assume that $C=C_0$ and $\Lambda_0 C =  C\Lambda$ [first line of \cref{fig: imps approximations}(b)].
Now, in order to contract down the left environment and the iMPS unit cell, we have to bring the latter to \textit{left} canonical form by inserting the Schmidt values $\Lambda^{-1}$ and $\tilde\Lambda$ on the left and right ends of the unit cell, respectively.
In all, the overlap $\braket{\tilde\Psi|\Psi}$ is reduced to $\tr(\tilde\Lambda^2D_0D^\dagger)$, where $\tilde\Lambda$ are the Schmidt values in the longer chain at the right end of the iMPS unit cell.
The optimal $D$ follows from the same Lemma as above.
The lowest possible deviation is given by
\begin{align}
    \big\Vert \ket\Psi - \ket{\tilde\Psi}\big\Vert^2 
    &= \tr\tilde\Lambda^2+\tr\tilde\Lambda^2 - \tr(\tilde\Lambda D_0D^\dagger\tilde\Lambda) - \tr(\tilde\Lambda DD_0^\dagger\tilde\Lambda) \nonumber\\
    &= \tr\left[(\tilde\Lambda D - \tilde\Lambda D_0)\times\mathrm{H.c.}\right] + \tr\left[\tilde\Lambda (\unity-D_0D_0^\dagger)\tilde\Lambda \right].
    \label{eq: iMPS error right}
\end{align}
\code{temfpy.iMPS.MPS_to_iMPS()} returns estimates of both error terms separately for both sides of the unit cell.
\end{appendix}





\bibliography{fermion_MPS,software,attila_everything,bib,cft,sl_kagome}


\end{document}